\documentclass[fleqn,twoside]{article}
\usepackage{gc,amsfonts,amssymb,cite}

\def\Gun{\mbox{m$^3$ kg$^{-1}$ s$^{-2}$}}
\def\year{{\rm year}}
\def\R{{\mathbb R}}
\def\p{\d}
\newcommand{\desc}[1]{\begin{description} \itemsep -2pt
    #1\end{description}}
\def\itbf#1{\emph{\textbf{#1}}}

\heads {Vitaly N. Melnikov}
 {Variations of constants as a test of gravity, cosmology and unified models. (Review)}

\begin{document}
\twocolumn[



\Title {Variations of constants as a test of gravity,\yy
    cosmology and unified models}

\Author{Vitaly N. Melnikov\foom 1}
    {Centre for Gravitation and Fundamental Metrology, VNIIMS,
        46 Ozyornaya St., Moscow, 119361, Russia;\\
     Institute of Gravitation and Cosmology, Peoples' Friendship University
        of Russia\\  \quad 6 Miklukho-Maklaya St., Moscow 117198, Russia}


\Abstract
{Gravitation as a fundamental interaction that governs all phenomena at
 large and very small scales, but still not well understood at a quantum
 level, is a cardinal missing link in unification of all physical
 interactions. Discovery of the present acceleration of the Universe, the
 dark matter and dark energy problems are also a great challenge to modern
 physics, which may lead to a new revolution in it. Integrable
 multidimensional models of gravitation and cosmology make up one of the
 proper approaches to studying basic issues and strong field objects, the
 early and present Universe and black hole (BH) physics in particular
 \cite{Mel2,Mel,Mel3}. Our main results within this approach are described
 for both cosmology and BH physics. Problems of the absolute $G$
 measurements and its possible time and range variations are reflections of
 the unification problem.

$\quad$ The choice, nature, classification and precision of determination of
 the fundamental physical constants as well as their role in a transition,
 expected in 2011, to new definitions of the main SI units, supposed to be
 based on fundamental physical constants and stable quantum phenomena, are
 described. The problem of temporal variations of constants is also
 discussed, temporal and spatial variations of $G$ in particular. A need for
 further absolute measurements of $G$, its possible range and time
 variations is pointed out. The multipurpose space project SEE is briefly
 described, aimed at measuring $G$ and its stability in space and time, with
 precision 3-4 orders better than at present. It may answer many important
 questions posed by gravitation, cosmology and unified theories. A project
 of a laboratory experiment for testing possible deviations from the
 Newton law is also presented.}

] 
\email 1 {melnikov@phys.msu.ru, melnikov@vniims.ru}

\section{Introduction}

  Studies in the previous century in the field of gravitation were mainly
  devoted to theoretical studies and experimental verification of general
  relativity (GR) and alternative theories of gravity with a strong
  stress laid on relations between macro- and micro-world phenomena or, in
  other words, between classical gravitation and quantum physics. Very
  intensive investigations in these fields were done in Russia by M.A.
  Markov, K.P. Staniukovich, Ya.B. Zeldovich, A.D. Sakharov and their
  colleagues starting from mid-60s. As a motivation there were: the
  existence of singularities in cosmology and black hole physics, the role
  of gravity at large and very small (Planckian) scales, attempts to create
  a quantum theory of gravity as for other physical fields, problem of
  possible variations of fundamental physical constants etc. A lot of work
  has been done in such areas as \cite{3}:
\desc{
\item[$-$] exact solutions with different fields as sources in GR,

\item[$-$] particlelike solutions with a gravitational field,

\item[$-$] quantum field theory in a classical gravitational background,

\item[$-$] quantum cosmology with fields (e.g., scalar), with the
    cosmological constant etc.,

\item[$-$] self-consistent treatment of quantum effects in cosmology,

\item[$-$] development of alternative theories of gravity:
    scalar-tensor, gauge, with torsion, bimetric etc.,

\item[$-$] possible variations of fundamental physical constants
    \cite{4,5,6,IM1,BIM1}.
    }

  As our main results of this period, one can mention \cite {3} the first
  quantum cosmological model with a cosmological constant (creation from
  nothing) (1972); the first classical cosmological models for a conformal
  scalar field (1968) and quantum cosmological models with minimal and
  conformal scalar fields (1971), the first nonsingular cosmological model
  with spontaneous symmetry breaking of a nonlinear conformal scalar field
  (1978-79), exact solutions for nonlinear electrodynamics, including the
  Born-Infeld one, the first exact solution for dilaton-type interaction
  with an electromagnetic field in GR, the first nonsingular particlelike
  purely field solution with gravity (1979). One should also mention the
  conclusion that only $G$ may vary with respect to atomic system of
  measurements, or in the Jordan-Brans-Dicke frame (1978).

  Since all attempts to quantize general relativity in a usual manner failed
  and it was proved that it is nonrenormalizable, it became clear that a
  promising trend goes along the lines of unification of all physical
  interactions which has started in the 70s. About this time, experimental
  studies of gravity in strong fields and gravitational waves started,
  giving a powerful speed-up in theoretical studies of such objects as
  pulsars, black holes, QSO's, active galactic nuclei, the early Universe
  etc., which continue now.

  But nowadays, when we think of the most important lines of future
  developments in physics, we can foresee that gravity will be essential not
  only by itself, but as a missing cardinal link of some theory unifying all
  existing physical interactions: gravitational, weak, strong and
  electromagnetic ones. Even in experimental activities, some crucial next
  generation gravitational experiments verifying predictions of unified
  schemes will be important.  Among them are:  MICROSCOPE and STEP ---
  testing the cornerstone Equivalence Principle, SEE --- testing the
  inverse square law (or new non-Newtonian interactions), testing possible
  time variations of the Newtonian constant $G$, measurement of the absolute
  value of $G$ with unprecedented accuracy \cite{SanD,Team}. All these
  experiments became tests of not only Gravity itself, but unified models of
  physical interactions as well. Of course, the gravitational-wave problem,
  verification of torsional, rotational (GPB), 2nd order and strong field
  effects remain important as well.

  We can also predict that the studies of gravity itself and within the
  unified models will give, in the next century and millennium, even more
  applications to our everyday life, as the electromagnetic theory gave us in
  the 20th century after very abstract fundamental investigations of Faraday,
  Maxwell, Poincar\'e, Einstein and others, who had never dreamed of such
  enormous applications of their works.

  Another very important feature which may be envisaged is an increasing
  role of fundamental physics studies, gravitation, cosmology and
  astrophysics in particular, in space experiments \cite{Solv}. Unique
  micro-gravity environments and the modern technology outbreak give nearly
  a perfect place for gravitational experiments which suffer a lot on Earth
  from its relatively strong gravitational field and gravitational fields of
  nearby objects due to the fact that there is no way of screening gravity.

  In the development of relativistic gravitation and dynamical cosmology
  after A. Einstein and A. Friedmann, we may notice three distinct stages:
  first, studies of models with matter sources in the form of a perfect
  fluid, as was originally done by Einstein and Friedmann. Second, studies
  of models with sources as different physical fields, starting from the
  electromagnetic and scalar ones, in both classical and quantum setting
  (see \cite{3}).  And third, which is really topical now, application of
  ideas and results of unified models for treating fundamental problems of
  cosmology and black hole physics, especially in high energy regimes and
  for explanation of the greatest challenge to modern physics, explaining
  the present acceleration of the Universe,the so-called dark energy
  problem. Multidimensional gravitational models play an essential role in
  the latter approach.

  The necessity of studying multidimensional models of gravitation and
  cosmology \cite{Mel2,Mel} is motivated by several reasons. First, the main
  trend of modern physics is unification of all known fundamental physical
  interactions: electromagnetic, weak, strong and gravitational ones. In
  the recent decades, there has been a significant progress in unifying weak
  and electromagnetic interactions and some more modest achievements in GUT,
  supersymmetric, string and superstring theories.

  Now, theories with membranes, $p$-branes and more vague M-theory are being
  created and studied. Having no definite successful theory of unification
  now, it is desirable to study the common features of these theories and
  their applications to solving basic problems of modern gravity and
  cosmology.  Moreover, if we really believe in unified theories, the early
  stages of the Universe evolution and black hole physics, as unique superhigh
  energy regions, and possibly even the low energy stage, when we observe the
  present acceleration, are the most proper and natural arena for them.

  Second, multidimensional gravitational models, as well as scalar-tensor
  theories of gravity, are theoretical frameworks for describing possible
  temporal and range variations of fundamental physical constants
  \cite{3,4,5,6}. These ideas have originated from the earlier papers by E.
  Milne (1935) and P. Dirac (1937) on relationships between the phenomena of
  micro- and macro-worlds, and up till now they are under thorough study both
  theoretically and experimentally. The possible discovery of the fine
  structure constant variations is now under critical further investigation.

  Lastly, applying multidimensional gravitational models to basic problems
  of modern cosmology and black hole physics, we hope to find answers to
  such long-standing problems as singular or nonsingular initial states,
  creation of the Universe, creation of matter and its entropy, the
  cosmological constant and coincidence problem, origin of inflation and
  specific scalar fields which may be necessary for its realization, the
  isotropization and graceful exit problems, stability and nature of
  fundamental constants \cite{4,Solv,KM-G}, the possible number of extra
  dimensions, their stable compactification, new revolutionary data on
  present acceleration of the Universe, dark matter and dark energy etc.

  Bearing in mind that multidimensional gravitational models are certain
  generalizations of GR which is tested reliably for weak fields up to 0.0001
  and partly in strong fields (binary pulsars), it is quite natural to
  inquire about their possible observational or experimental windows.  From
  what we already know, among these windows are:

\desc{
\item[$-$] possible deviations from the Newton and Coulomb laws, or new
interactions,

\item[$-$] possible variations of the effective gravitational constant with
a time rate smaller than the Hubble one,

\item[$-$] possible existence of monopole modes in gravitational waves,

\item[$-$] different behaviour of strong field objects, such as
multidimensional black holes, wormholes and $p$-branes,

\item[$-$] standard cosmological tests,

\item[$-$] possible nonconservation of energy in strong field objects and
accelerators if the brane-world ideas about gravity in the bulk turn out to
be true etc.
}

  Since modern cosmology has already become a unique laboratory for testing
  the standard unified models of physical interactions at energies far
  beyond the level of existing and future manmade accelerators and other
  installations on Earth, there exists a possibility of using
  cosmological and astrophysical data for discriminating between future
  unified schemes. Data on possible time variations or deviations from the
  Newton law as a new important test should also contribute to the unified
  theory choice.

  As there exist no accepted unified models, in our approach
  \cite{Mel2,Mel,IMJ2,toprew} we have adopted simple (but general from
  the viewpoint of the number of dimensions) models, based on
  multidimensional Einstein equations with or without sources of
  different nature:
\desc{
\item[$-$] the cosmological constant,

\item[$-$] perfect and viscous fluids,

\item[$-$] scalar and electromagnetic fields,

\item[$-$] their possible interactions,

\item[$-$] dilaton and moduli fields with or without potentials,

\item[$-$] fields of antisymmetric forms (related to $p$-branes) etc.
  }
  The main objective of our programme was and is to obtain exact
  self-consistent solutions (integrable models) for these systems and then
  to analyze them in cosmological, spherically and axially symmetric cases.
  In our view, this is a natural and most reliable way of studying highly
  nonlinear systems. It is done mainly within Riemannian geometry. Some
  simple models in integrable Weyl geometry and with torsion were studied
  as well. In many cases, we tried to single out models which do not
  contradict the available experimental or observational data on variations
  of $G$. In some cases we have used the methods worked out for arbitrary
  dimensions in studying 4D models as well.

  As our model \cite{Mel2,Mel}, we use $n$ Einstein spaces of constant
  curvature with sources as $(m+1)$-component perfect fluid (or form fields,
  $…$), cosmological or spherically symmetric metrics, in manifolds obtained
  as a direct product of factor spaces of arbitrary dimensions. Then, in the
  harmonic time guage, we show that the Einstein multidimensional equations
  are equivalent to Lagrange equations with a nondiagonal (in general)
  minisuperspace metric and some exponential potential. After possible
  diagonalization of this metric, we perform reduction to sigma-model and
  Toda-like systems and further to Liouville, Abel, generalized Emden-Fowler
  equations etc. and try to find exact solutions. We suppose that the
  behaviour of extra spaces is the following: they can be constant or
  dynamically compactified (e.g., toroidally), or large, but with barriers,
  walls etc.

  So, we have been realizing the programme in arbitrary dimensions (from
  1988) \cite{Mel2,Mel,Mel3,IMJ2,toprew}.

\textbf{In cosmology}:\\
  we have obtained exact general solutions of multidimensional Einstein
  equations with sources:
\desc{
\item[$-$] $\Lambda$, $\Lambda$ + scalar field (e.g. nonsingular, dynamically
   compactified, inflationary);

\item[$-$] perfect fluid (PF), PF + scalar field (e.g. nonsingular,
   inflationary solutions);

\item[$-$] viscous fluid (e.g. nonsingular, generation of mass and entropy,
   quintessence and coincidence in a 2-component model);

\item[$-$] stochastic behaviour near the singularity, billiards in Lobachevsky
   space, that $D=11$ is critical and $\varphi$ destroys billiards (1994).
   }
For all the above cases, Ricci-flat solutions were obtained for any $n$,
and, in addition, solutions with curvature in one factor space; with
curvatures in 2 factor spaces, solutions are known only for total $D=10, 11$;
\desc{
\item[$-$] fields: scalars, dilatons, forms of arbitrary rank (1998) ---
   inflationary solutions, $\Lambda$ generation by forms (p-branes)
   \cite{Cosm};

\item[$-$] the first billiards for dilaton-forms (p-branes) interaction (1999);

\item[$-$] quantum systems (solutions of the WDW equation \cite{BIMZ}) for all
  the above cases where classical solutions were obtained;

\item[$-$] dilatonic fields with potentials, billiard behaviour for them as
well.
    }
 For many of these integrable models, we have calculated the time variation
 of the effective gravitational constant and compared it with the present
 experimental bounds, which allowed choosing particular models or singling
 out some classes of solutions.

\textbf{Solutions depending on r in any dimensions}:
 \desc{
\item[$-$] generalized Schwarzschild and Tangherlini (BHs are singled out),
   solutions with a minimally coupled scalar field $\varphi$ (no BHs);

\item[$-$] generalized Reissner-Nordstr\"om (BHs are also singled out), the
   same plus $\varphi$ (no BHs);

\item[$-$] multitemporal solutions;

\item[$-$] for dilaton-like interaction of $\varphi$ and electromagnetic
   fields (BHs exist only in a special case);

\item[$-$] stability studies for the above solutions (only BH ones are stable);

\item[$-$] the same for dilaton-forms interaction (p-branes); stability was
   found only in some cases, e.g., for a single form.
}
The PPN parameters for most of the models were calculated.

\textbf{Theory of experiments:}

  Space and laboratory experiments aimed at testing a possible violation of
  the Newton law and raising the precision of the absolute value of the
  Newton constant $G$ determination were suggested and worked out.

\section{Multidimensional Models}

  The history of the multidimensional approach begins with the well-known
  papers of T.K. Kaluza and O. Klein on 5-dimensional theories, which
  aroused an interest in studies of multidimensional gravity. These ideas
  were continued by P. Jordan who suggested to consider the more general
  case $g_{55}\ne \const $ leading to a theory with an additional scalar
  field.  They were, in some sense, a source of inspiration for C. Brans and
  R.H.  Dicke in their well-known work on a scalar-tensor gravitational
  theory.  After their work, a lot of investigations have been performed
  with material or fundamental scalar fields, both conformal and
  non-conformal (see details in \cite{3}).

  A revival of ideas of many dimensions started in the 70s and is continuing
  now, entirely due to the development of unified theories. In the 70s, an
  interest in multidimensional gravitational models was mainly stimulated by
  (i) the ideas of gauge theories leading to a non-Abelian generalization of
  the Kaluza-Klein approach and (ii) by supergravitational theories. In the
  80s, the supergravitational theories were ``replaced'' by superstring
  models. Now, it is heated by expectations connected with the overall
  M-theory. In all these theories, 4-dimensional gravitational models with
  extra fields were obtained from some multidimensional model by dimensional
  reduction based on decomposition of the manifold as
\beq
    M = M^4\times  M_{\rm int},
\eeq
  where $M^4$ is a 4-dimensional manifold and $M_{\rm int}$ is some
  internal manifold (mostly considered to be compact).

  The earlier papers on multidimensional gravity and cosmology dealt with
  multidimensional Einstein equations and with a block-diagonal cosmological
  or spherically symmetric metric defined on the manifold $M= \R \times
  M_0\times \dots \times M_n$ of the form
\beq
    g=-dt\otimes dt+\sum_{r=0}^n a_r^2(t) g^r,
\eeq
  where $(M_r,g^r)$ are Einstein spaces, $g^r$ is a metric on $M_r, \
  r=0,\dots,n$. In some of them, simple scalar fields and a cosmological
  constant were also included \cite{BIMZ}.

  Such models are usually reduced to pseudo-Euclidean Toda-like systems with
  the Lagrangian
\beq
    L=\frac12G_{ij}\dot x^i\dot x^j-\sum_{k=1}^mA_k\e^{u_i^k x^i}
\eeq
  and the zero-energy constraint $E=0$.

  It should be noted that pseudo-Euclidean Toda-like systems are not yet
  well studied. There exists a special class of equations of state that
  gives rise to Euclidean Toda models \cite{GIM}.

  Cosmological solutions are closely related to solutions with spherical
  symmetry \cite{IME}.  The first multidimensional generalization of such
  type was considered by D. Kramer and rediscovered by A.I. Legkii, D.J.
  Gross and M.J. Perry (and also by Davidson and Owen). Moreover, the scheme
  of obtaining the latter is very similar to the cosmological approach
  \cite{Mel2,IMi}. We continued these investigations in detail. In
  \cite{BrI}, the Schwarzschild solution was generalized to the case of $n$
  internal Ricci-flat spaces, and it was shown that a black hole
  configuration takes place when the scale factors of internal spaces are
  constants. It was also shown there that a minimally coupled scalar field
  is incompatible with the existence of black holes. In \cite{FIM2}, an
  analogous generalization of the Tangherlini solution was obtained, and an
  investigation of singularities was performed in \cite{IMB}. These
  solutions were also generalized to the electrovacuum case with and
  without a scalar field \cite{FIM3,IM8,BM}.  Here, it was also proved that
  BHs exist only when the scalar field is switched off. Deviations from the
  Newton and Coulomb laws were obtained, depending on the mass, charge and
  number of dimensions. In \cite{BM}, spherically symmetric solutions were
  obtained for a system of scalar and electromagnetic fields with a
  dilaton-type interaction, and also deviations from the Coulomb law were
  calculated, depending on the charge, mass, number of dimensions and the
  dilatonic coupling.  Multidimensional dilatonic black holes were singled
  out. A theorem was proved in \cite{BM} that ``cuts'' all non-black-hole
  configurations as being unstable under even monopole perturbations. In
  \cite{IM13}, the extremely charged dilatonic black hole solution was
  generalized to multi-centre (Majumdar-Papapetrou) configurations when the
  cosmological constant is non-zero.

  We note that for $D =4$ the pioneering Majumdar-Papapetrou-type
  solutions with a conformal scalar field and an electromagnetic
  field were considered in \cite{Br}.

  At present, there exists a special interest in the so-called M-theory etc.
  These theories are ``super-membrane'' analogues of the superstring models
  in $D=11$ and higher dimensions. The low-energy limit of these theories
  leads to multidimensional models with p-branes.

\subsection*{Exact solutions with ``branes''}

  In our papers, several classes of the exact solutions were considered for
  the multidimensional gravitational model governed by the Lagrangian
\bearr
   {\cal L}  = R[g]- 2\Lambda - h_{\alpha\beta}
   g^{MN}\d_{M}\varphi^\alpha\d_{N}\varphi^\beta
\nnn \inch
   -\sum_{a}\frac{1}{n_a!}\exp(2\lambda_{a \alpha}\varphi^\alpha)
   (F^a)^2,
\ear
  Here $g$ is a metric, $F^a = d A^a$ are forms of ranks $n_a$,
  $\varphi^\alpha$ are scalar fields, and $\Lambda$ is a cosmological
  constant  (the matrix $h_{\alpha\beta}$ is invertible).

\medskip\noi
{\bf Supergravities.}
  For certain field contents with distinguished values of the total
  dimension $D$, ranks $n_a$, dilatonic couplings $\lambda_{a}$  and
  $\Lambda = 0$, such Lagrangians appear as ``truncated'' bosonic sectors
  (i.e. without Chern-Simons terms) of certain supergravitational theories
  or a low-energy limit of superstring models \cite{CJS,GrSW}.  For $D=11$
  supergravity \cite{CJS} (considered now as a low-energy limit of a
  conjectured $M$-theory \cite{Wit}), we have a metric and a 4-form in the
  bosonic sector. For $D = 10$, one may consider type I supergravity with
  a metric, a scalar field and a 3-form; type IIA supergravity, with bosonic
  fields of type I supergravity, called the Neveu-Schwarz-Neveu-Schwarz
  (NS-NS) sector, and additionally a 2-form and a 4-form Ramond-Ramond (R-R)
  sector; type IIB supergravity with bosonic fields of type I supergravity
  (the NS-NS sector) and additionally a 1-form, a 3-form and a (self-dual)
  5-form (the R-R sector). It is now believed that all five string theories
  (I, IIA, IIB and two heterotic ones with gauge groups $G = E_8 \times E_8$
  and ${\rm Spin(32)}/Z_2)$ \cite{GrSW} as well as 11-dimensional
  supergravity \cite{CJS} are limiting cases of M-theory. All these theories
  are conjectured to be related by a set of duality transformations: $S-$,
  $T-$ (and more general $U-$) dualities.

  It was proposed earlier that the IIB string may have originate in a
  12D theory, known as F-theory (Vafa). A low energy effective (bosonic)
  Lagrangian for F-theory was also suggested.  The field content of this
  12-dimensional field model is the following: a metric, a scalar field
  (with a negative kinetic term), a 4-form and a 5-form. In our work
  \cite{IMJ}, a chain of so-called  $B_D$-models in dimensions $D = 11, 12,
  \ldots$ was suggested. The $B_D$-model contains $l =D-11$ scalar fields
  with negative kinetic terms (i.e. the so-called ``phantom'' fields)
  coupled to $(l+1)$ different forms of ranks $4, \ldots, 4 + l$.  These
  models were constructed using $p$-brane intersection rules that will be
  discussed below. For $D=11$ ($l=0$), the $B_D$-model coincides with the
  truncated  bosonic sector of $D=11$ supergravity. For $D=12$ $(l=1)$, it
  coincides with the truncated $D=12$ model. We have conjectured in
  \cite{IMJ} that these $B_D$-models for $D > 12$ may correspond to
  low-energy limits of some unknown $F_D$-theories (analogues of $M-$ and
  $F$-theories).

\medskip\noi
  {\bf Description of the models.}
  In our review \cite{toprew}, certain classes of $p$-brane solutions to
  field equations corresponding to the Lagrangian (4), obtained by us
  earlier, were presented.

  These solutions have block-diagonal metrics
  defined on a $D$-dimensional product manifold, i.e.,
\beq
     g= \e^{2\gamma} g^0  + \sum_{i=1}^{n} \e^{2\phi^i} g^i, \quad\
        M_0  \times M_{1} \times \ldots \times M_{n},
\eeq
  where $g^0$  is  a metric on $M_0$ (our space) and $g^i$ are fixed
  Ricci-flat (or Einstein) metrics on $M_i$ (internal spaces, $i >0$). The
  moduli $\gamma,\ \phi^i$ and the scalar fields  $\varphi^{\alpha}$ are
  functions on $M_0$, and fields of forms are also governed by several
  scalar functions on $M_0$. Any $F^a$ is supposed to be a sum  of  monoms,
  corresponding to electric or magnetic $p$-branes ($p$-dimensional
  analogues of membranes), i.e., the so-called composite $p$-brane ansatz is
  considered \cite{IM11,IM12}. (In the non-composite case we have no more
  than one monomial for each $F^a$.) $p=0$ corresponds to a particle, $p=1$ to
  a string, $p=2$ to a membrane etc. The $p$-brane world volume (world line
  for $p=0$, world surface for $p=1$ etc.) is isomorphic to some product of
  internal manifolds: $M_I = M_{i_1} \times \ldots \times M_{i_k}$ where $1
  \leq i_1 < \ldots < i_k \leq n$ and has the dimension $p + 1 = d_{i_1} +
  \ldots + d_{i_k} = d(I)$, where $I = \{i_1, \ldots, i_k \}$ is a
  multi-index describing the location of the $p$-brane, and $d_i = \dim M_i$.
  Any $p$-brane is described by the triplet ($p$-brane index) $s = (a,v,I)$,
  where $a$ is the colour index labelling the form $F^a$, $v = e(lectric),
  m(agnetic)$ and $I$ is the multi-index defined above. For the electric and
  magnetic branes corresponding to a form $F^a$, the world volume dimensions
  are $d(I) = n_a-1$ and $d(I) = D-n_a-1$, respectively. A sum of these
  dimensions is $D - 2$. For $D =11$ supergravity we get $d(I) =3$ and $d(I)
  = 6$, corresponding to an electric $M2$-brane and a magnetic $M5$-brane.

\medskip\noi
 {\bf Sigma-model representation. }
  In  our paper \cite{IMC}, the model under consideration was reduced to
  a gravitating self-interacting sigma-model with certain constraints. The
  sigma-model representation for the non-composite electric case was
  obtained earlier in \cite{IM11,IM12}, for the electric composite case see
  also \cite{IMR}).

  The $\sigma$-model Lagrangian, obtained from (4), has the form  \cite{IMC}
\bearr
  {\cal L}_{\sigma } =
    R[g^0]- \hat G_{AB} g^{0\mu\nu}\p_\mu\sigma^A\p_\nu\sigma^B
\nnn \cm
    -\sum_{s}\eps_s \exp(-2U^s)g^{0\mu\nu} \p_\mu\Phi^s\p_\nu\Phi^s -2V,
\ear
  where $(\sigma^A)=(\phi^i,\varphi^\alpha)$, $V$ is a potential, $(\hat
  G_{AB})$ are components of the (truncated) target space metric, $\eps_s =
  \pm 1$,
\[
  U^s =   U_A^s \sigma^A =  \sum_{i \in I_s} d_i \phi^i -
        \chi_s \lambda_{a_s \alpha} \varphi^{\alpha}
\]
  are linear functions, $\Phi^s$ are scalar functions on $M_0$
  (corresponding to the forms), and $s = (a_s, v_s, I_s)$. The parameter
  $\chi_s = +1$ for the electric brane ($v_s =e$) and $\chi_s = -1$ for the
  magnetic one ($v_s =m$).

  The pure gravitational sector of the sigma model was considered earlier in
  our paper  \cite{IME}. For $p$-brane applications, $g^0$ is Euclidean,
  $(\hat G_{AB})$ is positive-definite (for $d_0 > 2$), and $\eps_s= -1$ if
  pseudo-Euclidean (electric and magnetic) $p$-branes in a pseudo-Euclidean
  space-time are considered. The sigma model (6) may also be considered for
  the pseudo-Euclidean metric $g^0$ of signature $(-,+, \ldots, +)$ (e.g. in
  studies of gravitational waves). In this case, for a positive-definite
  matrix $(\hat G_{AB})$ and $\eps_s= 1$, we get non-negative kinetic
  energy terms.

\medskip\noi
  {\bf The brane $U$-vectors.}
  The co-vectors $U^s$ play a key role in studying the integrability of the
  field equations \cite{IMC,IMBl} and possible existence of stochastic
  behaviour near the singularity, see our paper \cite{IMb1}. An important
  mathematical characteristic here is the matrix of scalar products
  $(U^s,U^{s'}) =\hat G^{AB} U_A^s U_B^{s'}$, where $(\hat G^{AB}) = (\hat
  G_{AB})^{-1}$. The scalar products for the co-vectors $U^s$  were
  calculated in \cite{IMC} (for the electric case see \cite{IM11,IM12,IMR}):
\bearr
    (U^s,U^{s'})=d(I_s\cap I_{s'})+\frac{d(I_s)d(I_{s'})}{2-D}
\nnn \inch
 + \chi_s\chi_{s'}\lambda_{a_s\alpha} \lambda_{a_{s'}\beta} h^{\alpha \beta},
\earn
  where $(h^{\alpha\beta})=(h_{\alpha\beta})^{-1}$; $s=(a_s,\ v_s,\ I_s)$,
  $s'=(a_{s'},\ v_{s'},\ I_{s'})$. They depend on brane intersections (the
  first term), the brane world-volume dimensions and the total dimension $D$
  (the second term), the scalar products of dilatonic coupling vectors and
  the electro-magnetic types of branes (the third term).  As will be shown
  below, the so-called ``intersections rules''(i.e., relations for
  $d(I_s\cap I_{s'})$) are determined by the scalar products of $U^s$-vectors.

\subsection*{Cosmological and spherically symmetric solutions}

  A family of general cosmological-type $p$-brane solutions with $n$
  Ricci-flat internal spaces was considered in our paper \cite{IMJ1}, where
  a generalization to the case of $n-1$ Ricci-flat spaces and one Einstein
  space of non-zero curvature (say, $M_1$) was also obtained. These
  solutions are defined up to solutions to Toda-type equations  and may be
  obtained using the Lagrange dynamics following from our sigma-model
  approach \cite{IMJ}. The solutions from \cite{IMJ1} contain a subclass of
  spherically symmetric solutions (for $M_1 = S^{d_1}$). Special solutions
  with orthogonal and block-orthogonal \cite{bobs} sets of $U$-vectors were
  considered previously in our works \cite{IMJ} and \cite{IMJ2,IMJ1},
  respectively. (For the non-composite case, see \cite{GrIM,BGIM}) and
  references therein.)

\medskip\noi
  {\bf Toda solutions.}
  In \cite{IMJ}, a reduction of $p$-brane cosmological-type solutions to
  Toda-like systems was first performed. General classes of $p$-brane
  (cosmological and spherically symmetric) solutions related to Euclidean
  Toda lattices associated with Lie algebras (mainly ${\bf A_m}$, ${\bf
  C_m}$ ones) were obtained by us in \cite{GM1,GM2,GM3,IMp2,IMp3}.

  A class of space-like brane ($S$-brane) solutions  (related to Toda-type
  systems) with a product of Ricci-flat internal spaces and $S$-brane
  solutions with special orthogonal intersection rules were considered in
  \cite{IMSel3,IMKim-ac}, and solutions with accelerated expansion (e.g.
  with a power-law and exponential behaviour of scale factors) were singled
  out.

\medskip\noi
  {\bf Black-brane solutions.}
  In our papers \cite{IMp2,IMp3}, a family of spherically symmetric
  solutions was investigated, and a subclass of black-hole configurations
  related to Toda-type equations under certain asymptotical conditions
  was singled out. These black-hole solutions are governed by the functions
  $H_s(z) > 0$ defined on the interval $(0, (2 \mu)^{-1})$, where $\mu > 0$
  is the extremality parameter, and obey the set of differential equations
  (equivalent to Toda-type ones)
\[
 \frac{d}{dz} \left( \frac{(1 - 2\mu z)}{H_s} \frac{d}{dz} H_s \right) =
  \bar B_s \prod_{s' }  H_{s'}^{- A_{s s'}},
\]
 with the following boundary conditions:
\[      \nq
     {\bf (i)}  \ H_{s}((2\mu)^{-1}{-}0) = H_{s0} \in (0, + \infty); \ \
     {\bf (ii)}  \ H_{s}(+ 0) = 1,
\]
 $s \in S$. Here  $\bar B_s \neq 0$ and  $(A_{s s'})$ is a quasi-Cartan
 matrix.

 In Refs.\,\cite{IMp1,IMp2,IMp3}, the following hypothesis was put forward:
 the functions $H_s$ are polynomials when the intersection rules
\bearr
   d(I_{s}\cap I_{s'})=
     \frac{d(I_s)d(I_{s'})}{D - 2} - \chi_s\chi_{s'}\lambda_{a_s \alpha}
        \lambda_{a_{s'} \beta} h^{\alpha \beta}
\nnn \cm\cm
    +\frac12 (U^{s'},U^{s'}) A_{s s'},  \cm s \neq s'.
\earn
 correspond to semi-simple Lie algebras, i.e., when $(A_{s s'})$ is a
 Cartan matrix. Here,
\[
    (A_{ss'}) \equiv \left(\frac{2(U^s,U^{s'})}{(U^{s'},U^{s'})}\right),
\]
  $s,s'\in S$, is a quasi-Cartan matrix.

 This hypothesis was verified for Lie algebras: ${\bf A_m }$, ${\bf
 C_{m+1}}$, $m = 1,2, \ldots$, in \cite{IMp2,IMp3}. It was also confirmed
 by special  black-hole ``block orthogonal'' solutions considered earlier in
 \cite{IMJ2,CIM}.

 In our papers \cite{IMp1,IMp2,IMp3}, explicit formulae for the solution
 corresponding to the algebra ${\bf A_2}$ are presented. These formulae are
 illustrated by two examples of ${\bf A_2}$-dyon solutions:  a dyon in $D =
 11$ supergravity (with M2- and M5-branes intersecting at a point) and a
 Kaluza-Klein dyon. Extremal configurations (e.g., with a multi-black-hole
 extension) were also obtained.

 We note that special black-hole solutions with orthogonal $U$-vectors
 were considered in  \cite{BIM} (for the non-composite case) and \cite{IMJ}.
 These solutions have analogues in models with a multi-component
 perfect fluid \cite{IMSel1,IMSel2,DIMel}.

 The black-brane solution, corresponding to the Lie algebras ${\bf C_2}$ and
 ${\bf A_3}$ were obtained in \cite{GrIM3}.

 In \cite{BIM}, some propositions related to i) interconnection between the
 Hawking temperature and the singularity behaviour, and ii) multi-temporal
 configurations were proved.

 It should be noted that the polynomial structure was also found for the
 so-called flux-brane solutions which occur as generalizations of the
 well-known Melvin solution.

\medskip\noi
{\bf Cosmological models in diverse dimensions.}
  Scalar fields play an essential role in modern cosmology though the
  problem of their origin still exists. They are attributed to inflation
  models of the early Universe and to models describing the present stage of
  the accelerated expansion. There is no unique candidate potential of the
  minimally coupled scalar field. Typically a potential is taken as a sum of
  exponentials. Such potentials appear quite generically in a large class of
  theories: Kaluza-Klein models, supergravity and string/M-theories.

  A single exponential potential was extensively studied in the
  Friedmann-Robertson-Walker (FRW) 4D model containing both a minimally
  coupled scalar field  and a perfect fluid with a linear barotropic
  equation of state. The attention was mainly focussed on the qualitative
  behaviour of solutions, stability of exceptional solutions to curvature
  and shear perturbations and their possible applications within the known
  cosmological scenarios such as inflation and scaling (``tracking''). In
  particular, it was found by phase plane analysis  that for ``flat''
  positive potentials there exists a unique late-time attractor in the form
  of a scalar-dominated solution. It is stable within homogeneous and
  isotropic models with non-zero spatial curvature with respect to spatial
  curvature perturbations and provides power-law inflation. For
  ``intermediate'' positive potentials, a unique late-time attractor is
  the scaling solution where the scalar field ``mimics'' a perfect fluid,
  adopting its equation of state. The energy density of the scalar field
  scales with that of the perfect fluid. Using our methods for
  multidimensional cosmology, the problem of integrability by quadratures of
  the model in 4 dimensions was also studied. Four classes of general
  solutions, when the parameter characterizing the steepness of the
  potential and the barotropic parameter obey some relations, were found
  \cite {DGavMel2}. For the case of multiple exponential potential of the
  scalar field plus dust, an integrable model in 4D was obtained in \cite
  {GavMelAbd}.

  As to scalar fields with a multiple exponential potential in any
  dimensions, it is not yet well studied, although a wide class of exact
  solutions were obtained in our papers \cite{IMjhep,AlIvMel}. In our
  recent work \cite{DIMbill}, the behaviour of this system near the
  singularity was studied using a billiard approach suggested earlier in our
  papers \cite{IMBill95,IMb1}. A number of S-brane solutions were found in
  \cite{IMSel3,IMKim-ac}. See details for 2-component D-dimensional
  integrable models in Refs.\,\cite{MelGav,GavMel0,GavMel1}).

  Quite a different model with a dilaton, branes and a cosmological
  constant, with static internal spaces, was investigated in \cite{Cosm},
  where possible generation of the effective cosmological constant by branes
  was demonstrated. A model with variable equations of state was found in
  \cite{AlGavMel}, with acceleration in our space and compactification
  of internal spaces.

\medskip\noi
  {\bf Cosmological models with time variations of $G$.}
  As was mentioned before, cosmological models in scalar-tensor and
  multidimensional theories are frameworks for describing possible time
  variations of fundamental physical constants due to scalar fields present
  explicitly in STT and present initially or/and generated by extra
  dimensions in multidimensional approach. In \cite{2}, we have obtained
  solutions for a system of conformal scalar and gravitational fields in
  4D and calculated the present possible relative variation of $G$ at the
  level of less than $10^{-12}/$year. Later, in the framework of a
  multidimensional model with a perfect fluid and 2 factor spaces (our 3D
  space could be open, closed or flat) and an internal 6D Ricci-flat space,
  we obtained the same limit for such variation of $G$ \cite{BIM1}.

  We have also estimated the possible variations of the gravitational
  constant $G$ in the framework of a generalized (Bergmann-Wagoner-Nordtvedt)
  scalar-tensor theory of gravity on the basis of the field equations,
  without using their special solutions. Specific estimates were essentially
  related to the values of other cosmological parameters (the Hubble and
  acceleration parameters, the dark matter density etc.), but the values of
  $\dot G/G$ compatible with modern observations $10^{-12}/$year
  \cite{BMN-G} were not exceeded.

  In \cite{MIWaseda}, we continued the studies of models in arbitrary
  dimensions and obtained relations for $\dot G$ in a multidimensional model
  with Ricci-flat internal space and a multi-component perfect fluid. A
  two-component example (dust + 5-brane) was considered. It was shown that
  $\dot G/G$ is less than $10^{-12}$/year. Expressions for $\dot G$ were
  also considered in a multidimensional model with an Einstein internal
  space and a multicomponent perfect fluid \cite{DIKM-G}. In the case of
  two factor spaces with non-zero curvatures without matter, a mechanism for
  prediction of small $\dot G$ was suggested.  The result was compared with
  exact (1+3+6)-dimensional solutions which we obtained earlier.

  A multidimensional cosmological model describing the dynamics of $n+1$
  Ricci-flat factor-spaces $M_i$ in the presence of a one-component
  anisotropic fluid was considered in \cite{AIKM}. The pressures in all
  spaces were supposed to be proportional to the density: $p_{i} = w_i
  \rho,\ i = 0,...,n$. Solutions with an accelerated power-law expansion of
  our 3-space $M_0$ and small enough variation of the gravitational constant
  $G$ were found. These solutions exist for two branches of the parameter
  $w_0$. The first branch describes superstiff matter with $w_0 > 1$,
  the second one may contain phantom matter with $w_0 < - 1$, e.g., when
  $G$ grows with time, so this branch can describe not the present epoch
  but rather earlier stages.

  Similar exact solutions, but nonsingular and with an exponential behaviour
  of the scale factors, were considered in \cite{IKMN} for the same
  multidimensional cosmological model describing the dynamics of $n+1$
  Ricci-flat factor spaces $M_i$ in the presence of a one-component perfect
  fluid. Solutions with accelerated exponential expansion of our 3-space
  $M_0$ and small variation of $G$ were also found.

  Exact S-brane solutions with two electric branes and two phantom
  scalar fields on the manifold
\beq
    M = \R_+ \times \R \times M_2 \times
                M_3 \times M_4 \times M_{5}.
\eeq
 were obtained and studied in \cite{AIMAG}. We obtained asymptotic
 accelerated expansion of our 3D factor space and variations of $G$ obeying
 the present experimental constraints, $\dot G/G \lesssim 10^{-12}$/year.

 A D-dimensional cosmological model with several scalar fields and an
 antisymmetric $(p+2)$-form was also considered \cite{DIMParal}. For
 dimensions $D = 4m+1 = 5, 9, 13, ...$ and $p = 2m-1 = 1, 3, 5, ...$, we
 obtained a family of new cosmological type solutions with a $4m$-dimensional
 oriented Ricci-flat submanifold $N$ of Euclidean signature. These
 solutions are characterized by a self-dual or anti-self-dual parallel
 charge density form $Q$ of rank $2m$ defined on $N$. The (sub)manifold $N$
 may be chosen to be either K\"ahler or hyper-K\"ahler, or an 8-dimensional
 manifold of $Spin(7)$ holonomy. A generalization of these solutions to a
 chain of extra (marginal) Ricci-flat factor spaces was also presented.
 Solutions with accelerated expansion of extra factor spaces were singled
 out. Certain examples of new solutions for IIA supergravity and for a chain
 of $B_D$-models in dimensions $D = 14, 15, ...$ were considered.

\medskip\noi
 {\bf Spherically symmetric solutions, black holes and PPN parameters}.
  In \cite{IMC}, it was shown that, after dimensional reduction on the
  manifold $M_0\times M_1\times\dots\times M_n$ and when the composite
  $p$-brane ansatz is considered, the problem is reduced to the gravitating
  self-interacting $\sigma$ model with certain constraints. For electric
  $p$-branes, see also \cite{IM12,IMR} (in \cite{IMR}, the composite
  electric case was considered). This representation may be considered as a
  powerful tool for obtaining different solutions with intersecting
  $p$-branes. In \cite{IMC,IMBl}, Majumdar-Papapetrou type solutions were
  obtained (for the non-composite electric case see \cite{IM12} and for the
  composite electric case see \cite{IMR}). These solutions correspond to
  Ricci-flat $(M_i,g^i)$, $i=1,\dots,n$ and were generalized to the case of
  Einstein internal spaces \cite{IMC}.  These solutions take place when
  certain {\em orthogonality relations\/} (on the couplings parameters,
  brane dimensions and the total dimension) are imposed. In this situation,
  a class of cosmological and spherically symmetric solutions was obtained
  \cite{IMJ}. Solutions with a horizon were considered in detail in
  \cite{BIM,IMJ}.

  It should be noted that multidimensional and multi-temporal
  generalizations of the Schwarz\-schild and Tan\-gherlini solutions were
  considered in \cite{IM8,IM6I}, where the generalized Newton formulae in
  the multitemporal case were obtained.

  We have also calculated the Post-Newtonian Parameters $\beta$ and
  $\gamma$ (the Eddington parameters) for general spherically symmetric
  solutions and black holes in particular \cite{IMJ2}. These parameters,
  depending on $p$-brane charges, their world-volume dimensions, dilaton
  couplings and the total number of dimensions may be useful for physical
  applications.

  Some specific models in classical and quantum multidimensional cases with
  $p$-branes were analyzed in \cite{IMJ,GrIM}. Exact solutions for a system
  of scalar fields and fields of forms with dilatonic type interactions
  for {\em generalized intersection rules\/} were studied in \cite{IMp2},
  where the PPN parameters were also calculated. Other problems connected
  with observations were studied in \cite{BrMpn,BrMelconf}, and general
  properties of BHs and wormholes in a brane world in \cite{BrKim,BrDM}.

  {\em Stability\/} analysis for solutions with $p$-branes was carried out
  in \cite{BrM,STAB}. It was shown there that, for some simple $p$-brane
  systems, multidimensional black branes are stable under monopole
  perturbations while other (non-BH) spherically symmetric solutions turned
  out to be unstable.

  Below we mainly dwell upon some problems of fundamental physical
  constants, the gravitational constant in particular, upon the SEE and
  laboratory projects for measuring $G$ and its possible variations and
  briefly on some theoretical models with variations of the effective
  gravitational constant.

\section{Fundamental physical constants}

  {\bf 1.} In any physical theory we meet constants which characterize the
  stability properties of different types of matter: objects, processes,
  classes of processes and so on. Some of them cannot be calculated via other
  physical constants. These constants are important because they arise
  independently in different situations and have the same value, at any rate
  within accuracies we have gained nowadays. That is why such constants are
  called the fundamental physical constants (FPCs) \cite{3,Solv}. It is
  impossible to define strictly this notion. It is because the constants,
  mainly dimensional ones, are present in certain physical theories. In the
  process of scientific progress, some theories are replaced by more general
  ones with their own constants, and there arise relations between old and
  new constants. So, we cannot speak of an absolute choice of FPC, but
  rather only of a choice corresponding to the present state of the physical
  sciences.

  Really, before creation of the electro-weak interaction theory and some
  Grand Unification Models, the following {\em choice\/} of FPCs was
  considered:
\beq  \nq
    c,\ \hbar,\ \alpha,\ G_{F},\ g_s,\ m_p\ ({\rm or}\
                m_e),\ G,\ H,\ \rho,\ \Lambda,\ k,\ I,
\eeq
  where $\alpha$, $G_F$, $g_s$ and $G$ are constants of the electromagnetic,
  weak, strong and gravitational interactions, $H$, $\rho$ and $\Lambda$ are
  cosmological parameters (the Hubble constant, the mean density of the
  Universe and the osmological constant), $k$ and $I$ are the Boltzmann
  constant and the mechanical equivalent of heat which play the role of
  conversion factors between temperature on the one hand, energy and
  mechanical units on the other. After adoption (in 1983) of a new
  definition of the meter ($\lambda = ct$ or $\ell = ct$), this role is
  also partly played by the speed of light $c$. It is now also a conversion
  factor between units of time (frequency) and length, it is defined with
  absolute accuracy, i.e., zero uncertainty (with the new suggested
  definitions of basic units of the International System of Units (SI), such
  a role may also be played by $\hbar$ and $N_{A}$, where $N_{A}$ is
  the Avogadro number \cite{Mills} .

  Now, when the theory of electro-weak interactions has a firm experimental
  basis and we have some good models of strong interactions, a more
  preferable choice is as follows:
\bearr
    \hbar,\ (c),\ e,\ m_e,\ \theta_w,\ G_F,\ \theta_c,\
    \Lambda_{\rm QCD},\
\nnn \inch
    G,\ H,\ \rho,\ \Lambda,\ k,\ I
\ear
  and maybe the three Kobayashi-Maskawa angles $\theta_2$, $\theta_3$ and
  $\delta$. Here $\theta_w$ is the Weinberg angle, $\theta_c$ is the Cabbibo
  angle and $\Lambda_{\rm QCD}$ is the cut-off parameter of quantum
  chromodynamics.  Of course, if a theory of the four interactions known now
  will be created (M-theory or some other), then we will probably have
  another choice. As we see, the macroscopic constants remain the same,
  though in some unified models, i.e. in multidimensional ones, they may be
  related in some manner (see below). From the point of view of these
  unified models, the above-mentioned ones are low-energy constants.

  All these constants are known with different {\em accuracies}. The most
  precisely defined constant was and remains to be the speed of light $c$:
  its uncertainty was $10^{-10}$ while now it is defined with an absolute
  accuracy. The atomic constants $e$, $\hbar$, $m$ and others are
  determined with errors $10^{-6}\div 10^{-8}$, $G$ up to $10^{-4}$ or even
  worse, $\theta_w$ up to $10^{-3}$; the accuracy of $H$ is about a few
  per cent. Other cosmological parameters (FPCs): mean density estimations
  vary within 2 per cent; for $\Lambda$ we now have data that its
  corresponding energy density exceeds the matter density (0.7 vs. 0.3 of
  the total Universe mass). Here are some recent estimates from
  observational cosmology:
\desc{
\item[$-$] Total mean density : $0.98 <  \Omega_{\rm tot}  < 1.08$.

\item[$-$] Today's Hubble parameter: $H_0 = 0.72 \pm 0.07$.

\item[$-$] Dark energy density parameter: $\Omega_{DE} = 0.7$.

\item[$-$] For dark matter: $\Omega_{DM} = 0.26$.

\item[$-$] For baryonic matter: $\Omega_B = 0.04$.

\item[$-$] For radiation: $\Omega_R = 5 \times 10^{-5}$.

\item[$-$] Power spectrum index: $n = 0.970 \pm 0.023$.

\item[$-$] Equation of state coefficient: $w = p/ \rho < - 0.78$.
    }

  As to the {\em nature\/} of the FPCs, we can mention several approaches.
  One of the first hypotheses belongs to J.A. Wheeler: in each cycle of the
  Universe evolution, the FPCs arise anew along with physical laws which
  govern this evolution. Thus the nature of the FPC and physical laws are
  connected with the origin and evolution of our Universe.

  A less global approach to the nature of the dimensional constants suggests
  that they are needed to make physical relations dimensionless or they are
  measures of asymptotic states. Really, the speed of light appears in
  relativistic theories in factors like $v/c$, at the same time velocities of
  usual bodies are smaller than $c$, so it also plays the role of an
  asymptotic limit. The same sense have some other FPCs: $\hbar$ is the
  minimal quantum of action, $e$ is the minimal observable charge (if we do
  not take into account quarks which are not observable in free state) etc.

  Finally, FPCs or their combinations may be considered as natural scales
  determining the basic units. While earlier the basic units were chosen
  more or less arbitrarily, i.e., the second, metre and kilogram, now the
  first two are based on stable (quantum) phenomena. Their stability is
  believed to be ensured by the physical laws which include FPCs. There
  appeared similar suggestions for a new reproducible realization of a kg,
  to fix the values of $N_{A}$ or other constants, e.g. $\hbar$ \cite{Mills}.

  Another interesting problem, which is under discussion, is why the FPCs
  have values in a very narrow range necessary for supporting life
  (stability of atoms, stellar lifetime etc.). There exist several possible
  explanations. First, that it is a good luck, no matter how improbable is
  the set of FPCs. Second, that life may exist in other forms and with
  another FPC set, of which we do not know. Third, that all possibilities
  for FPC sets exist in some universe. And the last but not least: that
  there is some cosmic fine tuning of FPCs, some unknown physical processes
  bringing them to their present values in a long-time evolution, cycles etc.

  An exact knowledge of FPCs and precision measurements are necessary for
  testing the main physical theories, extension of our knowledge of nature
  and, in a long run, for practical applications of fundamental theories.
  Within this, certain theoretical problems arise:

\desc{
\item[1)] development of models for confrontation of theory with experiment
    in critical situations (i.e., for verification of GR, QED, QCD, GUT
    or other unified models);

\item[2)] setting limits on spacial and temporal variations of the FPCs. It
    is becoming especially important now, with the idea to introduce new
    basic units of the International System of Units (SI), based
    completely on fundamental physical constants.
    }
  Of course, raising the precision of their absolute values is always
  a permanent task.

  As to a {\em classification\/} of the FPCs, we can set them into four
  groups according to their generality:

\desc{
\item[1)] Universal constants, such as $\hbar$ which divides all phenomena
    into quantum and non-quantum ones (micro- and macro-worlds) and to a
    certain extent $c$, which divides all motions into relativistic and
    non-relativistic ones;

\item[2)] coupling constants like $\alpha$, $\theta_w$, $\Lambda_{QCD}$ and
    $G$;

\item[3)] constants of elementary constituencies of matter like $m_e$, $m_w$,
    $m_x$, etc., and

\item[4)] transformation multipliers such as $k$, $I$ and partly $c$
    (conversion from the second to the metre). Soon there may be more
    after modernization of SI in 2011 --- the values of $\hbar$ and
    $N_{A}$ may be fixed with zero uncertainty.
    }
  Of course, this division into classes is not absolute. Many constants move
  from one class to another. For example, $e$ was a charge of a particular
  object, the electron, class 3, then it became a characteristic of class 2
  (electromagnetic interaction, $\alpha = e^2/(\hbar c)$ in combination
  with $\hbar$ and $c$); the speed of light $c$ has been in nearly all
  classes: from 3 it moved into 1, then also into 4. Some of the constants
  ceased to be fundamental (i.e. densities, magnetic moments, etc.) as they
  are calculated via other FPCs.

  As to the {\em number} of FPCs, there are two opposite tendencies: the
  number of ``old'' FPCs is usually diminishing when a new, more general
  theory is created, but at the same time new fields of science arise, new
  processes are discovered in which new constants appear. So, in the long
  run, we may come to some minimal choice which is characterized by one or
  several FPCs, maybe connected with the so-called Planck parameters ---
  combinations of $c$, $\hbar$ and $G$ (the natural, or Planck system of
  units \cite{Solv,KM-G}):
\bear
    L = (\hbar G/c^3)^{1/2} \lal\sim 10^{-33}\
    {\rm cm},
\nn
    m_L = (c\hbar/2G)^{1/2} \lal \sim 10^{-5}\ {\rm g},
\nn
        \tau_L = L/c \lal\sim 10^{-43}\ {\rm s}.
\ear
  The role of these parameters is important since $m_L$ characterizes the
  energy of unification of the four known fundamental interactions: strong,
  weak, electromagnetic and gravitational ones, and $L$ is a scale where the
  classical notions of space-time lose their meaning. There are other ideas
  about the final number of FPC (2, 1, or none, L. Okun' et al.) Of course,
  everything will depend on a future unified theory.

\bigskip\noi
  {\bf 2.} The problem of the gravitational constant $G$ measurement and its
  stability is part of a rapidly developing field, called
  gravitational-relativistic metrology (GRM). It has appeared due to the
  growth of measurement technology precision, spread of measurements over
  large scales and a tendency to unification of fundamental physical
  interactions, where the main problems arise and are concentrated on the
  gravitational interaction. This was first formulated in \cite{6}.

  The main subjects of GRM are:
\desc{
\item[$-$] general-relativistic models on different astronomical scales:
    Earth, the Solar system, galaxies, cluster of galaxies, cosmology;

\item[$-$] time transfer, VLBI, space dynamics, relativistic astrometry
        etc. (pioneering works were done in Russia by Arifov and Kadyev and
        Brumberg in the 60s);

\item[$-$] development of generalized gravitational theories and unified
    models for testing their effects in experiments;

\item[$-$] measurement of fundamental physical constants, $G$ in particular,
    and their stability in space and time; MICROSCOPE, STEP, SEE...

\item[$-$] fundamental cosmological parameters as fundamental constants:
    cosmological models studies (quint\-essence, k-essence, phantom,
    multidimensional models), measurements and observations;
    projects PLANCK, ...

\item[$-$] gravitational waves (detectors, sources...); LIGO, VIRGO, TAMA,
    LISA, RADIOASTRON,...

\item[$-$] basic standards (clocks) and other modern precision devices
    (atomic and neutron interferometry, atomic force spectroscopy
    etc.) in fundamental gravitational experiments, especially in
    space for testing GR and other theories: rotational, torsional and
    second-order effects (need an uncertainty of $10^{-6}$ or better),
    e.g.  LAGEOS, Gravity Probe B, ASTROD, LATOR etc.
    }
  We are now at the level of $2.3 \cdot 10^{-5}$ in measuring the
  PPN-parameter $\gamma$ and $5 \cdot 10^{-4}$ for $\beta$; the Brans-Dicke
  parameter is $\omega > 40000$.

\medskip
The proposed future missions aimed at improving the accuracy of $\gamma$ are:
\begin{enumerate}                   \itemsep -2pt
\item GP-B (geodetic precession) --- $10^{-5}$.

\item Bepi-Colombo (retardation) --- $10^{-6}$.

\item GAIA (deflection) -- $(10^{-5} - 10^{-7})$.

\item ASTROD I (W.-T. Ni) (retardation) --- $10^{-7}$.

\item LATOR (Turyshev et al.) --- $10^{-8}$.

\item ASTROD (W.-T. Ni) --- $10^{-9}$.
\end{enumerate}

  There are three problems related to $G$, whose origin is mainly related to
  unified models predictions:
\desc {
\item[1)] absolute $G$ measurements,

\item[2)] possible time variations of $G$,

\item[3)] possible scale variations of $G$ --- non-Newtonian, or new
    interactions.
    }

\noi
  {\bf Absolute measurements of $G$}. There are many laboratory
  determinations of $G$ with errors of the order $10^{-3}$, and only 4 have
  been on the level of $10^{-4}$ in the 80s. They are given in Table 1 (in
  $10^{-11}$ \Gun).

\medskip \noi
    Table 1  \medskip

{\small \nqq\
\begin{tabular}{lll}
1. Facy and Pontikis, France & 1972  &  6,6714 $\pm$ 0.0006\\
2. Sagitov et al., Russia    & 1979  &  6,6745 $\pm$ 0.0008\\
3. Luther and Towler, USA    & 1982  &  6,6726 $\pm$ 0.0005\\
4. Karagioz, Russia          & 1988  &  6,6731 $\pm$ 0.0004\\
\end{tabular}     }
\medskip

  From this table it is evident that the first three experiments contradict
  each other (the results do not overlap within their uncertainties). And
  only the fourth experiment is in accord with the third one.

  The official CODATA value of 1986 was
\beq
    G = (6,67259 \pm 0.00085)\cdot 10^{-11}\ \Gun,
\eeq
  based on Luther and Towler's determination. But after precise measurements
  of $G$ by different groups, the situation became more vague.

  As one may see from the Cavendish conference data of 1998 \cite{MT99}, the
  results of 7 groups could agree with each other only on the level
  of $10^{-3}$. So, CODATA adopted in 1999
\beq
    G=(6.673 \pm 0.001) \cdot 10^{-11}\ \Gun.
\eeq

  The most recent and precise $G$ measurements do not agree, and some of
  them differ from the CODATA value of 1986. They are given in Table 2 (in
  $10^{-11}$ \Gun)

\begin{table*}
Table 2     \medskip

    \inch
\begin{tabular}{lll}
1. Gundlach and Merkowitz (USA) \cite{G}  & 2000 & $6,674215 \pm 0,000092$\\
2. Armstrong (New Zealand, MSL)        & 2002 & $6,6742 \pm 0,0007$    \\
3. Karagioz (Moscow, Russia)           & 2003 & $6.6729 \pm 0.0005$    \\
4. Luo Zhun (Wuhan, China)             & 1998 & $6,6699 \pm 0,0007$    \\
5. Quinn et al. (BIPM)  (1)            & 2001 & $6,6693 \pm 0,0009$    \\
 \phantom{Quinn et al. (BIPM)}\quad\ (2) & 2001 & $6,6689 \pm 0,0014$    \\
6. Schlamminger et al. (CH)            & 2002 & $6.7404 \pm 0.000033$  \\
\end{tabular}
\end{table*}

  But, from 2004 CODATA gives:
\beq
   G = 6.6742(10) \cdot 10^{-11} \Gun.
\eeq
  So we see that we are not too far (a little more than one order) from
  Cavendish who obtained a value of $G$ at the level of $10^{-2}$ two
  centuries ago. The situation with the measurement of the absolute value of
  $G$ is really different from that with atomic constants values and their
  uncertainties. This means that either the limit of terrestrial accuracies
  of determining $G$ has been reached or we have some new physics
  enterfering the measurement procedure \cite{6}. The former means that
  maybe we should turn to space experiments to measure $G$ \cite{Solv,Team},
  while the latter means that a more thorough study of theories generalizing
  Einstein's GR or unified theories is necessary.

  There also exist some satellite determinations of $G$ (namely,  $G\cdot
  M_{\rm Earth}$) at the level of $10^{-9}$ (so, could we know $G$ much
  better, our knowledge of masses of the Earth and other planets and
  consequently their models would be much better).  There are also several
  less precise geophysical determinations in mines at the level of
  $10^{-3}$, but they do not improve the situation.

  A precise knowledge of $G$ is necessary, above all, since it is a FPC;
  next, for evaluation of mass of the Earth, planets, their mean density
  and, finally, for construction of Earth models; for transition from
  mechanical to electromagnetic units and vice versa; for evaluation of other
  constants through relations between them obtained by unified theories; for
  finding new possible types of interactions and geophysical effects; for
  some practical applications like increasing of modern gradiometer
  precision, since they demand calibration by a gravitational field of a
  standard body depending on $G$: the high accuracy of their calibration
  ($10^{-5}$ - $10^{-6}$) requires the same accuracy of $G$.

\medskip\noi
  {\bf 3.} The knowledge of constants values has not only a fundamental but
  also a {\it metrological} meaning. The modern system of standards is
  mainly based on stable physical phenomena. So, the stability of constants
  plays a crucial role. And it may be even more important if new definitions
  of the units via fixed fundamental constants in the International System
  of Units (SI) will be introduced in 2011, as is suggested now.  Let us
  give some short historical references to the evolution of systems of units.

  Before the official adoption of the Metric Convention (1875) in 1799, two
  platinum basic standards (\emph{\textbf{metre and kilogram}}) were
  introduced and placed in the Archive of the French Republic, which started
  the Metric System.

  Gauss in 1832 created the first coherent system of units introducing the
  second and measured magnetic fields in terms of mechanical units: mm,
  gram and \emph{\textbf{second}}.

  In 1874, the British Association for Development of Science introduced the
  coherent system of units CGS based on Maxwell's and Thomson's suggestions
  of 1860: \itbf {centimetre ---  gram --- second}.

  In 1880, the British Association and International Congress of
  Electricians adopted practical units --- ohm, volt and \itbf{ampere}.

  In 1889, the First General Committee on Weights and Measures (BIPM)
  adopted the MKS System (metre, kilogram, second).

  In 1939, the MKSA system (MKS + ampere) was suggested (based on Georgi's
  suggestion of 1901) and officially adopted by BIPM in 1946.

  In 1954. \itbf{kelvin and candela} were introduced by BIPM, and the whole
  system of 6 basic units was called SI (International System of Units) in
  1960.

  In 1971, the \itbf{mole} was added to SI, so up till now {\bf SI has 7
  basic units: metre, kilogram, second, ampere, kelvin, candela and mole}.

  A new stage of SI evolution started in October 2005 when the International
  Committee for Weights and Measures (CIPM) adopted a recommendation on
  preparative steps towards redefining the

  \itbf{kilogram, ampere, kelvin and mole},\\
  so that these units be linked to exactly known values of fundamental
  constants. Mills et al. \cite{Mills} proposed that these four base units
  should be given new definitions linking them to exactly defined values of
  the Planck constant $h$, elementary charge $e$, the Boltzmann constant $k$
  and the Avogadro constant $N_A$, respectively. This would mean that six
  of the seven base units of the SI would be defined in terms of true
  invariants of nature. Not only would these four fundamental constants
  have exactly defined values, but also the uncertainties of many of the
  other fundamental constants of physics would be either eliminated or
  appreciably reduced. They suggested wordings for the four new definitions
  and presented a novel way to define the entire SI units explicitly using
  such definitions without making any distinction between base units and
  derived units:

    \itbf{The metre, unit of length}, is such that the {\bf speed of
    light in vacuum  $c$} is \itbf{exactly}

    299 792 458  metres  per  second.

  Such a definition is simple, concise and makes clear the fundamental
  constant to which the unit is linked and the exact value of that constant.

  If this general form were chosen, it would be appropriate to choose
  definitions of the same form for all seven base units.

  Thus, for the second, candela, kilogram, ampere, kelvin and mole we would
  have:

\medskip\noi
    \itbf{The second, unit of time}, is such that the ground state
    hyperfine splitting transition frequency of the caesium 133 atom is
    \itbf{exactly}

    9 192 631 770  hertz.

\medskip\noi
    \itbf{The candela, unit of luminous intensity in a given direction},
    is such that the spectral luminous efficacy of monochromatic
    radiation of frequency$540 \times 1012$ hertz is \itbf{exactly}

    683 lumens  per watt.

\medskip\noi
    \itbf{The kilogram, unit of mass,} is such that the \textbf{Planck
    constant} is \itbf{exactly}

    $6.626 069 3 \times 10^{-34}$  joule  second.

\medskip\noi
    \itbf{The ampere, unit of electric current,} is such that the
    elementary charge is \itbf{exactly}

    $1.602 176 53 \times 10^{-19}$ coulomb.

\medskip\noi
    \itbf{The kelvin, unit of thermodynamic temperature}, is such that
    the \textbf{Boltzmann constant} is \itbf{exactly}

    $1.380 650 5 \times 10^{-23}$ joule  per  kelvin.

\medskip\noi
    \itbf{The mole, unit of amount of substance of a specified elementary
    entity}, which may be an atom, molecule, ion, electron, any other
    particle or a specified group of such particles, is such that the
    \textbf{Avogadro constant} is \itbf{exactly}

    $6.022 141 5 \times 10^{23}$  per  mole.

\medskip
  Of course, there still remain a lot of problems to be solved before real
  introduction of these new definitions (raising the precision of absolute
  values of some constants, choosing the variant of {\bf kg} realization,
  testing the self-consistency of definitions, possible stability of
  constants, readiness of the world community to accept these changes
  (science, industry, trade, educational level etc.)

\medskip\noi
  {\bf 4.} {\em Time Variations of $G$}. As all physical laws were
  established and tested during the last 2--3 centuries in experiments on
  Earth and in the near space, i.e., at rather short space and time
  intervals as compared with the radius and age of the Universe, the
  possibility of slow {\em variations\/} of constants (i.e., with the rate
  of the evolution of the Universe or slower) cannot be excluded a priori.
  So, the assumption of absolute stability of constants is an extrapolation,
  and each time we must test it.

  The problem of FPC variations arose with the attempts to explain the
  relations between micro- and macro-world phenomena. Dirac was the first to
  introduce (in 1937) the so-called ``Large Number Hypothesis'' which
  relates some known very big (or very small) numbers with the dimensionless
  age of the Universe $T\sim 10^{40}$ (age of the Universe in seconds
  $10^{17}$, divided by the characteristic elementary particle time
  $10^{-23}$ seconds).  He suggested that the ratio of the gravitational to
  strong interaction strengths, $Gm_p^2/(\hbar c)\sim 10^{-40}$, is
  inversely proportional to the age of the Universe: $Gm_p^2/\hbar c\sim
  T^{-1}$. Then, as the age varies, some constants or their combinations
  must vary as well. The atomic constants seemed to Dirac to be more stable,
  so he chose variation of $G$ as $T^{-1}$.

  After the original {\it Dirac hypothesis}\/, some new ones appeared
  (Gamow, Teller, Landau, Terazawa, Staniu\-ko\-vich etc., see \cite{3,Solv}),
  as well as some generalized theories of gravitation admitting variations
  of an effective gravitational coupling. We can single out three stages in
  the development of this field:
\begin{enumerate}  \itemsep -2pt
\item
    Studies of theories and hypotheses with FPC variations,
    their predictions and confrontation with experiments (1937-1977).

\item
    Creation of theories admitting variations of an effective
    gravitational constant in a particular system of units, analyses of
    experimental and observational data within these theories
    \cite{1,3} (1977-present).

\item
    Analyses of FPC variations within unified models \cite{6,4,Mel2}
    (present).
\end{enumerate}

  In the development of the first stage from the analysis of the whole set
  of the then available astronomical, astrophysical, geophysical and
  laboratory data, a conclusion was made \cite{1,2} that variations of
  atomic constants were excluded at the level of $10^{-15}$ per year, but
  variations of the effective gravitational constant in the atomic system of
  units do not contradict the experimental data on the level of $10^{-12}
  \div 10^{-13} \year^{-1}$.  Moreover, in \cite{1,2,7}, a conception was
  worked out that variations of constants are not absolute but depend on the
  system of measurements (choice of standards, units and devices using this
  or that fundamental interaction). Each fundamental interaction, through
  dynamics described by the corresponding theory, defines the system of
  units and the corresponding system of basic standards, e.g., atomic and
  gravitational (ephemeris) seconds.

  Earlier reviews of some hypotheses on variations of FPCs and experimental
  tests can be found in \cite{3,4,6}.

\medskip\noi
  {\bf 5.} There are different astronomical, geophysical and
  laboratory data on possible FPC variations \cite{Solv}.

\medskip\noi
  {\bf Astrophysical data.} Here follow some recent ones.

  Comparing the data from absorption lines of atomic and molecular
  transition spectra in high-redshift QSO's, Varshalovich and Potekhin
  (Russia) \cite{VP} obtained for $z = 2.8 - 3.1$:
\beq
    | \dot \alpha /\alpha | \leq 1.6\cdot10^{-14} \ \year^{-1},
\eeq
  and Drinkwater et al. \cite{D}:
\beq
    | \dot \alpha /\alpha | \leq 10^{-15} \ \year^{-1}\mbox{ for } z = 0.25
\eeq
and
\beq
    |\dot \alpha/\alpha| \leq 5\cdot 10^{-16} \ \year^{-1}\mbox{ for } z=0.68
\eeq
  for a model with zero deceleration parameter and $H = 75 {\rm km}\cdot
  {\rm s}^{-1}\cdot {\rm Mpc}^{-1}$.

  A less precise conclusion was made on the basis of {\em geophysical data}.
  Indeed, according to F. Dyson (1972), $\beta$-decay of $Re_{187}$ to
  $Os_{187}$ give:
\beq
    | \dot {\alpha}/\alpha | \leq 5\cdot10^{-15} \ \year^{-1}.
\eeq

  We must point out that all astronomical and geophysical estimations are
  strongly model-dependent. So, of course, it is always desirable to have
  {\em laboratory tests\/} of FPC variations.
\desc {
\item
    [a)] Such a test was first done by a Russian group in the Committee for
    Standards (Kolosnitsyn, 1975). Comparing rates of two different types of
    clocks, one based on a Cs standard and another on a beam molecular
    generator, they found that
\beq
    | \dot {\alpha}/\alpha | \leq 10^{-10} \ \year^{-1}.
\eeq

\item [b)] More recent data were obtained by J. Prestage et al. \cite{JP} by
    comparing mercury and $H$-maser clocks. Their result is
\beq
    | \dot {\alpha}/\alpha | \leq 3.7\cdot 10^{-14} \ \year^{-1}.
\eeq
    }
  All these limits were placed on the fine structure constant variations.
  From the analysis of decay rates of $K_{40}$ and $Re_{187}$, a limit on
  possible variations of the weak interaction constant was obtained (see
  ab approach for variations of $\beta$, e.g., in \cite{8})
\beq
    | \dot {\beta}/\beta | \leq 10^{-10} \ \year^{-1}.
\eeq

  But the most stringent data on variations of strong ($G_S$), electromagnetic
  and week ($G_W$) interaction constants were obtained by A. Schlyachter
  (USSR) in 1976 from an analysis of the ancient natural nuclear reactor
  data in Gabon, Oklo, because the event took place $2\cdot 10^9$ years ago.
  They are as follows:
\bear
    | \dot {G}_s/G_s| \lal < 5\cdot10^{-19} \ \year^{-1},
\nn
    | \dot {\alpha}/\alpha | \lal < 10^{-17} \ \year^{-1},
\nn
    | \dot {G}_F/G_F| \lal < 2\cdot10^{-12} \ \year^{-1}.
\ear

  Some studies of the strong interaction constant and its dependance on
  transfered momenta may be found in \cite{KrKM}. A recent review
  on variations of $\alpha$ see in \cite{BKalpha}.

  There have appeared some data on a possible variation of $\alpha$ on the
  level of $10^{-16}$ at some $z$ \cite{Webb}. Other groups do not support
  these results. There also appeared data on possible variation of
  $m_{e}/m_{p}$ (Varshalovich et al.)  The problem may be that even if they
  are correct, all these results are mean values of variations at some epoch
  of the evolution of the Universe (a certain $z$ interval). In essence,
  variations may be different at different epochs (if they exist at all), and
  at the next stage observational data should be analyzed with the account of
  evolution of the corresponding (``true''?) cosmological models.

\medskip\noi
  {\bf 6.} Now we are still having no unified theory of all four interactions.
  So it is possible to construct systems of measurements based on any of
  these four interactions. But in practice it is now done on the basis of
  the mostly worked out theory, electrodynamics (more precisely, on QED).
  Of course, it may also be done on the basis of the gravitational
  interaction (as was partly the case earlier). Then, different units of
  basic physical quantities arise, based on dynamics of a particular
  interaction, i.e., the atomic (electromagnetic) second, defined via
  frequency of atomic transitions, or the gravitational second defined by the
  mean Earth motion around the Sun (the ephemeris time).

  It does not follow from anything that these two seconds are always
  synchronized in time and space. So, in principle, they may evolve relative
  to each other, for example, at the rate of the evolution of the Universe
  or at some slower rate.

  That is why, in general, variations of the gravitational constant are
  possible in the atomic system of units ($c$, $\hbar$, $m$ are constant,
  Jordan frame) and masses of all particles --- in the gravitational system
  of units ($G$, $\hbar$, $c$ are constant by definition, Einstein frame).
  In practice, we can test only the first variant since the modern basic
  standards are defined in the atomic system of measurements. Possible
  variations of the FPCs must be tested experimentally, but for this purpose
  it is necessary to have the corresponding theories admitting such
  variations and their certain effects.

  Mathematically, these systems of measurement may be realized as
  conformally related metric forms. Arbitrary conformal transformations give
  us a transition to an arbitrary system of measurements.

  We know that the scalar-tensor and multidimensional theories are the
  corresponding frameworks for these variations. So, one of the ways to
  describe variable gravitational coupling is the introduction of a {\em
  scalar field\/} as an additional variable of the gravitational interaction.
  It may be done by different means (e.g. Jordan, Brans-Dicke, Canuto and
  others). We have suggested a variant of gravitational theory with a
  conformal scalar field (Higgs-type field \cite{9,3}), where Einstein's
  GR may be considered as a result of spontaneous breaking of the conformal
  symmetry (Domokos, 1976) \cite{3}. In our variant, a spontaneous breaking
  of the global gauge invariance leads to a nonsingular cosmology \cite{16}.
  Besides, we may get variations of the effective gravitational constant in
  the atomic system of units when $m$, $c$, $\hbar$ are constant and
  variations of all masses in the gravitational system of units ($G$, $c$,
  $\hbar$ are constant). It is done on the basis of approximate \cite{10}
  and exact cosmological solutions with local inhomogenity \cite{14}.

  The effective gravitational constant is calculated using the equations of
  motion. Post-Newtonian expansion is also used in order to confront the
  theory with existing experimental data. Among the post-Newtonian
  parameters, the parameter $f$ describing variations of $G$ is included. It
  is defined as
\beq
        \frac{1}{GM}\frac{d(GM)}{dt} = fH.
\eeq
  According to Hellings' data \cite{18} from the Viking mission,
\beq
    \tilde{\gamma}-1 = (-1.2\pm 1.6)\cdot 10^3, \qquad
    f = (4\pm 8)\cdot 10^{-2}.
\eeq
  In the theory with a conformal Higgs field \cite{10,14}, we have obtained
  the following relation between $f$ and $\tilde\gamma$:
\beq
    f = 4(\tilde{\gamma}-1).
\eeq
  Using Hellings' data for $\tilde{\gamma}$, we can calculate $f$ in our
  variant and compare it with $f$ from \cite{18}. Then we get $f = (-9,6\pm
  12,8)\cdot 10^{-3}$, which agrees with (24) within its accuracy.

  We have used here only Hellings' data on variations of $G$. Other
  theoretical calculations in different models give the following
  predictions: less than $10^{-12}$ per year in multidimensional models
  \cite{BIM1,MIWaseda,M-G,DIKM-G}, less than $10^{-14}$ per year \cite{BMN-G}.

  But the situation with the experiment and observations is not so simple.
  Along with \cite{18}, there are some other data \cite{3,4,Solv}, but the
  most precise are:
\begin{enumerate}  \itemsep -2pt

\item Hellings' result
\beq
    | \dot {G}/G| < (2\pm 4)\cdot 10^{-12}\ \year^{-1}.
\eeq

\item  A result from nucleosythesis (Acceta et al., 1992):
\beq
    |\dot{G}/G|<(\pm 0.9)\cdot 10^{-12}\ \year^{-1}.
\eeq

\item  E.V. Pitjeva's result (Russia) \cite{P}, based on satellites
       and planets motion:
\beq
        |\dot{G}/G|<(0\pm 2)\cdot 10^{-12}\ \year^{-1}
\eeq

\item   Some new results from pulsars and Big Bang nucleosynthesis (BBN)
    at the level of $10^{-12}$ per year.

  There are also BBN data of Copi et al., 2003:
\beq
    -3 \cdot 10^{-13} < \dot{G}/G <  4 \cdot 10^{-13}.
\eeq
\end{enumerate}

  As to other experimental or observational data, the results are of
  different quality. The most reliable ones are based on lunar laser ranging
  (LLR) (Muller et al, 1993, Williams et al, 1996, Nordtvedt, 2003). They
  are not better than $10^{-12}$ per year. Here, once more we see that there
  is a need for corresponding theoretical and experinmental studies.
  Probably, future space missions like Earth SEE-satellite \cite{SanD, Team,
  Solv,KM-G} or missions to other planets and lunar laser ranging will be a
  decisive step in solving the problem of temporal variations of $G$ and
  determining the fates of different theories which predict them, since the
  greater is the time interval between successive masurements and, of course,
  the more precise they are, the more stringent results will be obtained.

  As we saw, different theoretical schemes lead to temporal variations of the
  effective gravitational constant:
\begin{enumerate}  \itemsep -2pt
\item
  Empirical models and theories of Dirac type, where $G$ is replaced with
  $G(t)$.

\item Numerous scalar-tensor theories of Jordan-Brans-Dicke type where $G$
  depends on the scalar field $\sigma (t)$.

\item Gravitational theories with a conformal scalar field arising
  in different approaches \cite{1,2,3,9,20}.

\item Multidimensional unified theories in which there are dilaton fields
  and effective scalar fields appearing in our 4-dimensional spacetime from
  additional dimensions \cite{19,Mel2}. They also may help in solving the
  problem of a variable cosmological constant changing from very large
  Planckian to present (0.7 of total energy) values.
\end{enumerate}

  As was shown in \cite{4,19,Mel2}, temporal variations of FPCs are connected
  with each other in {\em multidimensional models\/} of unification of
  interactions. So, experimental tests on $\dot {\alpha}/\alpha$ may at the
  same time be used for estimation of $\dot {G}/G$ and vice versa. Moreover,
  variations of $G$ are also related to the cosmological parameters $\rho$,
  $\Omega$ and $q$, which gives opportunities of raising the precision of
  their determination.

  Since FPC variations are closely connected with the behaviour of internal
  scale factors, it is a direct probe of properties of extra dimensions and
  the corresponding unified theories \cite{IM1,BIM1,Mel2}. From this point of
  view, it is an additional test of not only gravity and cosmology, but of
  unified theories of physical interactions as well.

\medskip\noi
{\bf 7.} {\bf Non-Newtonian interactions, or range variations of $G$}.
  Nearly all modified theories of gravity and unified theories also predict
  some deviations from the Newton law (inverse square law, ISL) or
  composition-dependent violations of the Equivalence Principle (EP) due to
  appearance of new possible massive particles (partners) \cite{4}.
  Experimental data exclude the existence of these particles on a very good
  level in nearly all ranges except less than the {\it millimetre\/} and
  {\em metres and hundreds of metres} ranges. Our recent analysis of
  experimental bounds and new limits on possible ISL violation using a new
  method and modern precession data from satellites, planets, binary pulsar
  and LLR data were obtained in \cite{KolMel}.

  In the Einstein theory $G$ is a true constant. But, if we think that $G$
  may vary with time, then, from a relativistic point of view, it may vary
  with distance as well. In GR massless gravitons are mediators of the
  gravitational interaction, they obey second-order differential equations
  and interact with matter with a constant strength $G$. If any of these
  requirements is violated, we come in general to deviations from the Newton
  law with range (or to a generalization of GR).

  In \cite{5} we have analyzed several classes of such theories:
\begin{enumerate}  \itemsep -2pt
\item
  Theories with massive gravitons like bimetric ones or theories with
    a $\Lambda$-term.
\item
  Theories with an effective gravitational constant like the general
  scalar-tensor ones.
\item
  Theories with torsion.
\item
  Theories with higher derivatives (4th-order equations etc.), where
  massive modes appear, leading to short-range additional forces.
\item
  More elaborated theories with other mediators in addition to gravitons
  (partners), like supergravity, superstrings, M-theory etc.
\item
  Theories with nonlinearities induced by any known physical interactions
    (Born-Infeld etc.)
\item
  Phenomenological models where a detailed mechanism of deviation is not
  known (fifth or other force).
\item
  Modifications of the Newton law at large ranges (MOND etc.), small
  acceleration at $a > a_{0}$ (Pioneer anomaly, etc.)
\end{enumerate}
  In all these theories, some effective or real masses appear leading
  to Yukawa-type (or power-law) deviations from the Newton law,
  characterized by the strength $\alpha$ and the range $\lambda$.

  There exist some model-dependent estimates of these forces. The most
  well-known one belongs to Scherk (1979) from supergravity where the
  graviton is accompanied by a spin-1 partner (graviphoton) leading to an
  additional repulsion. Other models were suggested by Moody and Wilczek
  (1984) --- introduction of a pseudo-scalar particle leading to an
  additional attraction between macro-bodies with the range $2\cdot10^{-4}$
  cm $< \lambda < 20$ cm and strength $\alpha$ from 1 to $10^{-10}$ in this
  range. Another supersymmetric model was elaborated by Fayet (1986, 1990),
  where a spin-1 partner of a massive graviton gives an additional repulsion
  in the range of the order $10^{3}$ km and $\alpha$ of the order $10^{-13}$.

  A scalar field to adjust $\Lambda$ was introduced by S. Weinberg in 1989,
  with a mass smaller than $10^{-3} {\rm eV}/c^{2}$, or a range greater than
  0.1 mm. One more variant was suggested by Peccei, Sola and Wetterich
  (1987), leading to additional attraction with a range smaller than 10 km.

  Some $p$-brane models (ADD, brane worlds) also predict non-Newtonian
  additional interactions of both Yukawa or power-law types, in particular,
  in the submillimetre range, which is intensively discussed nowadays
  \cite{KM-G,BKM-BW}. On PPN parameters for multidimensional models with
  $p$-branes see above, \sect 2.

\medskip\noi
  {The Pioneer anomaly.}
  A more serious evidence on a possible violation of Newton's law has come
  to us from space, namely, from data processing on the motion of the
  spacecrafts Pioneer 10 and 11, referring to length ranges of the order
  of or exceeding the size of the Solar system. The discovered anomalous
  (additional) acceleration is \cite{m7}
\[
       (8.60 \pm 1.34) \ten{-8}\ {\rm cm/s^2},
\]
  it acts on the spacecrafts and is directed towards the Sun. This
  acceleration is not explained by any known effects, bodies or influences
  related to the design of the spacecrafts themselves (leakage etc.), as was
  confirmed by independent calculations.

  Many different approaches have been analyzed both in the framework of
  standard theories and invoking new physics (Chongming Xu's talk at ICGA-7,
  Taiwan, November 2005), but none of them now seems to be sufficiently
  convincing and generally accepted. There are the following approaches
  using standard physics:
\desc{
\item[$-$]
    an unknown mass distribution in the Solar system (Kuiper's belt),
    interplanetary or interstellar dust, local effects due to the
    Universe expansion \cite{m6};
\item[$-$]
    employing the Schwarzschild solution with an expanding boundary
    \cite{m7,m8} etc.
    }

Among the approaches using new physics one can mention:
\desc {
\item[$-$]
     a variable cosmological constant \cite{m9};
\item[$-$]
     a variable gravitational constant \cite{m12};
\item[$-$]
     a new PPN theory connecting local scales with the cosmological
     expansion \cite{m10};
\item[$-$]
     the five-dimensional Kaluza-Klein (KK) theory with a time-variable
     fifth dimension and varying fundamental physical constants \cite{m11};
\item[$-$]
     Moffat's \cite{m13} non-symmetric gravitational theory;
\item[$-$]
     Milgrom's \cite{m14,m15} modified Newtonian dynamics (MOND);
\item[$-$]
     special scalar-tensor theories of gravity \cite{m16};
\item[$-$]
     approaches using some ideas of multidimensional theories \cite{m17};
\item[$-$]
     modified general relativity with a generalized stress-energy
     tensor \cite{m18} etc.
    }

  This Pioneer anomaly has caused new proposals of space missions with more
  precise experiments and a wide spectrum of research at the Solar system
  length range and beyond:
\desc {
\item[$-$]
   Cosmic Vision 2015-2025, suggested by the European Space Agency, and
\item[$-$]
   Pioneer Anomaly Explorer, suggested by NASA \cite{m19}.
    }

  So, we hope they can contribute a lot to our knowledge of gravity
  and unified models.

\medskip\noi
{\bf 8. Space project SEE and laboratory projects.}
  We have seen that there are three problems connected with $G$. There is
  a promising new multi-purpose space experiment SEE (Satellite Energy
  Exchange) \cite{SanD,Team} which addresses all these problems and may be
  more effective in solving them than other laboratory or space experiments.

  This experiment is based on a limited 3-body problem of celestial
  mechanics: small and large masses in a drag-free satellite and the Earth.
  Unique horse-shoe orbits, which are effectively one-dimensional, are used
  in it.

  The aims of the SEE-project are to measure the Inverse Square Law (ISL) and
  the Equivalence Principle (EP) at ranges of metres and the Earth radius,
  $\dot G$ and the absolute value of $G$ with unprecedented accuracies.

  We have studied many aspects of the SEE project \cite{Team, Solv}:
\begin{enumerate}  \itemsep -2pt
\item
   A wide range of trajectories with the aim of finding optimal ones:
    \desc{
    \item[$-$] circular in a spherical field;
    \item[$-$] the same plus the Earth's quadrupole modes;
    \item[$-$] elliptic with eccentricity less than 0.05.
        }
\item Estimations of other celestial bodies' influence.

\item Estimation of the sensitivity of trajectories to changes in $G$ and
      the Yukawa-type interaction strength parameter $\alpha$.

\item Modelling the measurement procedures for $G$ and $\alpha$ by different
   methods for different ranges and for different satellite altitudes:
       \desc{
    \item[$-$] optimal, 1500 km,
        \item[$-$] ISS free flying platform, 500 km and
        \item[$-$] a high orbit, 3000 km.
                   }
\item Estimations of some sources of error were made for:
    \desc{
    \item[$-$] radial oscillations of the shepherd's surface;
        \item[$-$] longitudeal oscillations of the capsule;
        \item[$-$] transversal oscillations of the capsule;
        \item[$-$] shepherd's non-sphericity;
        \item[$-$] limits on the quadrupole moment of the shepherd;
        \item[$-$] limits on admissible charges and time scales of charging by
        high energy particles etc.
         }
\item Error budgets for $G$, ${\dot G}$ and $G(r)$ were calculated.
\end{enumerate}
  On the basis of all these studies, the general conclusion was that
  realization of the SEE project may improve our knowledge of $G$, $\dot G$
  and $G(r)$ by 3-4 orders of magnitude.

  A laboratory experiment was also suggested in our paper \cite{KKM} to test
  the possible range variations of $G$. It is an experiment for possible
  detection of new forces, or test of the inverse square law, parameterized
  by a Yukawa-type potential with strength $\alpha$ and range $\lambda$. The
  installation comprises a ball with a spherical cavity whose centre is
  shifted with respect to the ball centre. The ball is placed on a
  turn-table being subject to uniform rotation. A torsion balance as a
  sensitive element is placed inside the cavity. A uniform gravitational
  field created inside the ball does not affect the balance, but any
  non-gravitational forces create a torque which acts periodically during
  the rotation of the ball. The spectrum of harmonics was calculated. It is
  shown that preferable to use is the first harmonic in the measurements.
  Sensitivity of the method was evaluated, which is limited by uncertainties
  due to manufacturing of the elements and temperature fluctuations of the
  sensitive element. It was shown that the sensitivity of the method
  suggested may be at the level of $10^{-10}$ in $\alpha$ in the range of
  $\lambda$ from  0.1 to $10^{7}$ m in the space of the Yukawa parameters
  $(\alpha, \lambda)$.

\Acknow {The author is grateful to Prof. J.-M. Alimi for hospitality during
his stay in Paris in March, 2007. The work was supported in part by the
Observatoire de Paris-Meudon and RFBR project 05-02-17478-a.}


\small
\begin{thebibliography}{199}

\bibitem{Mel2}
V.N. Melnikov, ``Multidimensional Classical and Quantum Cosmology and
Gravitation. Exact Solutions and Variations of Constants.'' CBPF-NF-051/93,
Rio de Janeiro, 1993; \\
V.N. Melnikov, in: ``Cosmology and Gravitation'', ed. M. Novello,
Editions Frontieres, Singapore, 1994, p. 147.

\bibitem{Mel}
V.N. Melnikov, ``Multidimensional Cosmology and  Gravitation'',
CBPF-MO-002/95, Rio de Janeiro, 1995, 210 p.; \\
V.N. Melnikov. In: {\it Cosmology and Gravitation. II}, ed. M.
Novello, Editions Frontieres, Singapore, 1996, p. 465.\\

\bibitem{Mel3}
V.N. Melnikov, ``Exact Solutions in Multidimensional Gravity and Cosmology
III." CBPF-MO-03/02,  Rio de Janeiro, 2002, 297 pp.

\bibitem{3}
K.P. Staniukovich and V.N. Melnikov, ``Hydrodynamics, Fields and Constants
in the Theory of Gravitation'', Energoatomizdat, Moscow, 1983, 256 pp. (in
Russian). See English translation of the first 5 sections in:\\
V.N. Melnikov, ``Fields and Constraints in the Theory of Gravitation'', CBPF
  MO-02/02, Rio de Janeiro, 2002, 145 pp.

\bibitem{4}
V.N. Melnikov, {\it Int. J. Theor. Phys.} {\bf 33}, 1569 (1994).

\bibitem{5}
V. de Sabbata, V.N. Melnikov and P.I. Pronin,
{\it Prog. Theor. Phys.} {\bf 88}, 623 (1992).

\bibitem{6}
V.N. Melnikov. In: ``Gravitational Measurements, Fundamental Metrology
and Constants'', eds. V. de Sabbata and V.N. Melnikov, Kluwer Academic
Publ., Dordtrecht, 1988, p. 283.

\bibitem{IM1}
V.D. Ivashchuk and V.N. Melnikov, {\it Nuovo Cim.} {\bf B 102}, 131 (1988).

\bibitem{BIM1}
K.A. Bronnikov, V.D. Ivashchuk and V.N. Melnikov,
{\it Nuovo  Cim.} {\bf B 102}, 209 (1988).

\bibitem{SanD}
 A. Sanders and  W. Deeds. {\it Phys. Rev.\/} {\bf D 46}, 480 (1992).

\bibitem {Team}
  A.D. Alexeev, K.A. Bronnikov, N.I. Kolosnitsyn, M.Yu. Konstantinov,
  V.N. Melnikov and A.G. Radynov. {\it Izmeritelnaya tekhnika,} 1993,
  No. 8, p. 6; No. 9, p. 3; No. 10, p. 6; 1994, No. 1, p. 3;
  {\it Int. J. Mod. Phys.\/} {\bf D 3}, 773 (1994).

  P.N. Antonyuk, K.A. Bronnikov and V.N. Melnikov, {\it Astron. Lett.\/}
  {\bf 20}, 59 (1994).

  K.A. Bronnikov, M.Yu. Konstantinov and V.N. Melnikov,
  \GC {2} 361 (1996); {\it Izmerit. Tekhnika\/} 1996, No. 5, p. 3;
  \GC {3} 293 (1997).

  A.D. Alexeev, V.N.Melnikov et al., {\it Izmerit. Tekhnika\/} 1997,
  No. 10, p. 3; \GC {5} 67 (1999).

  A.J. Sanders, V.N. Melnikov et al., \GC {3} 287 (1997);
   {\it Meas. Sci. Technol.} {\bf 10}, 514 (1999); \CQG {17} 2331 (2000).

  V.N. Melnikov, {\it Science in Russia\/}, 2000, No. 6, p. 3.

\bibitem {Solv}
  V.N. Melnikov, ``Gravity as a Key Problem of the Millennium'',
    Proc. 2000 NASA/JPL Conference on Fundamental Physics in
    Microgravity, NASA Document D-21522, 2001, p. 4.1--4.17, Solvang,
    CA, USA.

\bibitem {KM-G}
  S.A. Kononogov and V.N. Melnikov,
     ``The fundamental physical constants, the gravitational constant,
       and the SEE space experiment project''.
       {\it Izm. Tekhnika\/} {\bf 6}, 1 (2005);
    {\it Measurement Techniques\/} {\bf 48}, 6, 521 (2005).

\bibitem{IMJ2}
   V.D. Ivashchuk and V.N. Melnikov, ``Multidimensional cosmological and
       spherically symmetric solutions with intersecting p-branes''.
    {\it In:\/} Lecture Notes in Physics, Vol. 537,
    ``Mathematical and Quantum Aspects of Relativity and Cosmology'',
    Proceedings of the Second Samos Meeting on Cosmology, Geometry and
    Relativity held at Pythagoreon, Samos, Greece, 1998,
    eds.: S. Cotsakis, G.W. Gibbons. Berlin, Springer, 2000;
    gr-qc/9901001.

\bibitem{toprew}
     V.D. Ivashchuk and V.N. Melnikov,  ``Exact solutions in
     multidimensional gravity with antisymmetric forms'',
     topical review, {\it Class. Quantum Grav.\/} {\bf 18},
     R82-R157 (2001); hep-th/0110274.

\bibitem{BIMZ}
   U. Bleyer, V.D. Ivashchuk, V.N. Melnikov and A.I. Zhuk,
   ``Multidimensional classical and quantum wormholes in models with a
    cosmological constant'', {\it Nucl. Phys.} {\bf B 429}, 117  (1994);
    gr-qc/9405020.

 \bibitem{GIM}
    V.R. Gavrilov, V.D. Ivashchuk and  V.N. Melnikov,
    {\it J. Math. Phys. } {\bf 36}, 5829 (1995).

\bibitem{IME}
    V.D. Ivashchuk and V.N. Melnikov, ``Multidimensional gravity with
    Einstein internal spaces'',  \GC {2} 177 (1996);hep-th/9612054.

\bibitem{IMi}
    V.D. Ivashchuk and V.N. Melnikov, ``Multidimensional cosmology with
    $m$-component perfect fluid'', {\it Int. J. Mod. Phys.\/} {\bf D 3}, 795
       (1994); gr-qc/9403063.

\bibitem{BrI}
   K.A. Bronnikov and V.D. Ivashchuk, {\it in:\/} Abstr. 8th Sov. Grav.
    Conf., Erevan, EGU, 1988, p. 156.

\bibitem{FIM2}
   S.B. Fadeev, V.D. Ivashchuk and V.N. Melnikov, {\it Phys. Lett. }
    {\bf A 161}, 98 (1991).

\bibitem{FIM3}
   S.B. Fadeev, V.D. Ivashchuk and V.N. Melnikov, {\it Chinese Phys. Lett. }
    {\bf 8}, 439 (1991).

\bibitem{IM8}
   V.D. Ivashchuk and V.N. Melnikov,
    {\it Class. Quantum Grav.}, {\bf 11}, 1793 (1994).

\bibitem{BM}
   K.A. Bronnikov and V.N. Melnikov, {\it Annals of Physics (N.Y.) } {\bf
    239}, 40 (1995).

\bibitem{IMB}
   V.D. Ivashchuk and V.N. Melnikov,
    ``On singular solutions in multidimensional gravity'',
    \GC {1} 204 (1996); hep-th/9612089.

\bibitem{IM13}
   V.D. Ivashchuk and V.N. Melnikov, ``Extremal dilatonic black holes in
   string-like model with cosmological term'', {\it Phys. Lett. } {\bf B
    384}, 58 (1996).

\bibitem{Br}
   N.M. Bocharova, K.A. Bronnikov and V.N. Melnikov,
    {\it Vestnik MGU (Moscow Univ.)}, {\bf 6}, 706 (1970);
    {\it Moscow Univ. Phys. Bull.\/} {\bf 25}, 6, 80 (1970)
    --- the first MP-type solution with a conformal scalar field; \\
   K.A. Bronnikov, ``The Reissner-Nordstr\"om problem in the presence of
    an electromagnetic field'', Prepr. ITF-72-20P, Kiev, 1972;\\
   K.A. Bronnikov and V.N. Melnikov,  {\it in:\/}
    {\it Problems of Theory of Gravitation and Elementary Particles},
    {\bf 5}, 80 (1974) (in Russian) --- the first MP-type solution with
    conformal scalar and electromagnetic fields.

\bibitem{CJS}
   E. Cremmer, B. Julia and J. Scherk, {\it Phys. Lett.\/} {\bf B 76},
    409 (1978).

\bibitem{GrSW}
   M.B. Green, J.H. Schwarz and E. Witten,  ``Superstring Theory''
    (Cambridge University Press., Cambridge, 1987).

\bibitem{Wit}
   E. Witten, ``String theory dynamics in various dimensions'',
    {\it Nucl. Phys.\/} {\bf B 443}, 85 (1995); hep-th/9503124.

\bibitem{IMJ}
   V.D. Ivashchuk and V.N. Melnikov,
    ``Multidimensional classical and quantum cosmology with intersecting
    $p$-branes'', {\it J. Math. Phys.}, {\bf 39}, 2866 (1998);
    hep-th/9708157.

\bibitem{IM11}
    V.D. Ivashchuk and V.N. Melnikov, ``Intersecting p-brane solutions in
    multidimensional gravity and M-Theory'',
    \GC {2} 297 (1996); hep-th/9612089.

\bibitem{IM12}
   V.D. Ivashchuk and V.N. Melnikov,
    {\it Phys. Lett. }  {\bf B 403}, 23-30 (1997).

\bibitem{IMC}
   V.D. Ivashchuk and V.N. Melnikov,
   ``Sigma-model for the generalized  composite p-branes'',
    {\it Class. Quantum Grav.\/} {\bf 14}, 3001 (1997); hep-th/9705036;
    Corrigenda; {\it ibid.,} {\bf 15}, 3941 (1998).

\bibitem{IMR}
    V.D. Ivashchuk, V.N. Melnikov and M. Rainer, ``Multidimensional
    sigma-models with composite electric p-branes'', \GC {4} 73 (1998);
     gr-qc/9705005.

\bibitem{BIM}
   K.A. Bronnikov, V.D. Ivashchuk and V.N. Melnikov,
   ``The Reissner-Nordstr\"om problem for intersecting electric and magnetic
    p-branes'', \GC {3} 203 (1997); gr-qc/9710054.

\bibitem{GrIM}
    M.A. Grebeniuk, V.D. Ivashchuk and V.N. Melnikov,
    ``Integrable multidimensional quantum cosmology  for intersecting
    p-branes'', \GC {3} 243 (1997); gr-qc/9708031.

\bibitem{BGIM}
    K.A. Bronnikov, M.A. Grebeniuk, V.D. Ivashchuk and V.N. Melnikov,
    ``Integrable multidimensional cosmology for intersecting p-branes'',
    \GC {3} 105 (1997).

\bibitem{bobs}
    K.A. Bronnikov, ``Block-orthogonal brane systems, black holes and
    wormholes'', \GC {4} 49 (1998); hep-th/9710207.

\bibitem{IMBl}
    V.D. Ivashchuk and V.N. Melnikov,
    ``Madjumdar-Papapetrou type solutions in sigma-model and
    intersecting p-branes'', {\it Class. Quantum Grav.\/} {\bf 16}, 849
    (1999); hep-th/9802121.

 \bibitem{GM1}
   V.R. Gavrilov and V.N. Melnikov,
   ``Toda chains with type $A_m$  Lie algebra for multidimensional classical
   cosmology with intersecting $p$-branes''. {\it In:\/}  Proceedings of
   the International seminar ``Curent Topics in Mathematical Cosmology'',
    (Potsdam, Germany, 30 March -- 4 April 1998), Eds. M. Rainer and
    H.-J. Schmidt, World Scientific, 1998,  p. 310; hep-th/9807004.

\bibitem{IMJ1}
   V.D. Ivashchuk and V.N. Melnikov. ``Cosmological and spherically
   symmetric solutions with intersecting p-branes'', {\it J. Math. Phys.\/}
   {\bf 40} (12), 6558-6576 (1999).

\bibitem{CIM}
   S. Cotsakis, V.D. Ivashchuk and V.N. Melnikov,
   ``P-brane black holes and post-Newtonian approximation'',
    \GC {5}, 52 (1999); gr-qc/9902148.

\bibitem{GM2}
   V.R. Gavrilov and V.N. Melnikov, ``Toda chains  associated with Lie
   algebras  $A_m$ in multidimensional gravitation and cosmology with
   intersecting $p$-branes'', {\it Theor. Math. Phys.\/} {\bf 123} (3),
    374 (2000).

\bibitem{GM3}
   S. Cotsakis, V.R. Gavrilov and V.N. Melnikov,
   ``Spherically symmetric solutions for p-brane models associated with
   Lie algebras'', \GC {6}, 66 (2000).

\bibitem{IMp1}
   V.D. Ivashchuk and V.N. Melnikov, ``P-brane black holes for general
   intersections'', \GC {5} 313 (1999); gr-qc/0002085.

\bibitem{IMp2}
   V.D. Ivashchuk and V.N. Melnikov, ``Black hole p-brane solutions for
   general intersection rules'', \GC {6} 27 (2000);  hep-th/9910041.

\bibitem{IMp3}
   V.D. Ivashchuk and V.N. Melnikov. ``Toda p-brane black holes
   and polynomials related to Lie algebras'', \CQG {17} 2073 (2000);
   math-ph/0002048.

\bibitem{BrM}
   K.A. Bronnikov and V.N. Melnikov, ``p-brane black holes as
    stability islands'', {\it Nucl. Phys.\/} {\bf B 584}, 436-458
    (2000); hep-th/0002200.

\bibitem{IMb1}
   V.D. Ivashchuk and V.N. Melnikov, ``Billiard representation for
   multidimensional cosmology with intersecting p-branes near the
   singularity''  {\it J. Math. Phys.\/} {\bf 41}, 6341 (2000);
    hep-th/9904077.

\bibitem{IMSel1}
    V.D. Ivashchuk, V.N. Melnikov and A.B. Selivanov,
    ``Multidimensional black hole solution in a model
    with an anisotropic fluid'', \GC {7} 308 (2001); gr-qc/0205103.

\bibitem{GrIM3}
     M.A. Grebeniuk, V.D. Ivashchuk and V.N. Melnikov,
     ``Black-brane solution for $A_3$ algebra'',
     {\it Phys. Lett. }, {\bf B 543}, 98 (2002);  hep-th/0208083.

\bibitem{IMjhep}
     V.D. Ivashchuk, V.N. Melnikov and A.B. Selivanov,
     ``Cosmological solutions in multidimensional model with multiple
     exponential potential'',  {\it JHEP} 0309, 059 (2003); hep-th/0309027.

\bibitem{IMSel2}
  V.D. Ivashchuk, V.N. Melnikov and A.B. Selivanov,
  ``Simulation of intersecting black brane solutions by multicomponent
  anisotropic fluid'', \GC {9}, 50 (2003); hep-th/0211247.

\bibitem{DIMel}
  H. Dehnen, V.D. Ivashchuk and V.N. Melnikov,
  ``On black hole  solutions  in a model with an anisotropic fluid'',
  \GC {9} 153 (2003);  gr-qc/0211049.

\bibitem{IMBill95}
  V.D. Ivashchuk and V.N. Melnikov
  ``Billiard representation for multidimensional cosmology with
  multicomponent perfect fluid near the singularity'',
  \CQG {12} (3) 809 (1995).

\bibitem{DIMbill}
  H. Dehnen, V.D. Ivashchuk and V.N. Melnikov,
  ``Billiard representation for multidimensional multi-scalar
    cosmological model with exponential potentials'',
   \GRG {36} 1563 (2004); hep-th/0312317.

\bibitem{IMSel3}
   V.D. Ivashchuk, V.N. Melnikov and A.B. Selivanov,
   ``Composite S-brane solutions on product of Ricci-flat spaces'',
    \GRG {36} 1593 (2004);  hep-th/0404113.

\bibitem{IMKim-ac}
   V.D. Ivashchuk, V.N. Melnikov and S.-W. Kim,
   ``S-brane solutions with acceleration in models with forms and multiple
   exponential potential'', \GC {10} 141--148 (2004); hep-th/0405009.

\bibitem{BrMpn}
   K.A. Bronnikov and  V.N. Melnikov.
   ``On observational predictions from multidimensional gravity'',
    \GRG {33} 1549--1578 (2001); gr-qc/0103079.

\bibitem{BrDM}
    K.A. Bronnikov, H. Dehnen and V.N. Melnikov,
      ``General class of black holes in a brane world'',
      {\it Phys. Rev.\/} {\bf D 68} 024025 (2003); gr-qc/0304068.

\bibitem{BrKim}
      K.A. Bronnikov and Sug-Won Kim,
      ``Possible wormholes in a brane world'',
      \PRD {67} 064027 (2003); gr-qc/0212112.

\bibitem{BrMelconf}
      K.A. Bronnikov and V.N. Melnikov,
      ``Conformal frames and D-dimensional gravity'',
      {\it in:\/} Proceedings of the 18th Course of the School on Cosmology
      and Gravitation: The Gravitational Constant.  Generalized
      Gravitational Theories and Experiments (30 April-10 May 2003, Erice).
      Ed. by G.T. Gillies, V.N. Melnikov and V. de Sabbata, Kluwer
      Acad. Publ., 2004, pp. 39--64; gr-qc/0310112.

\bibitem{GavMel0}
        V.R. Gavrilov and V.N. Melnikov,
        ``D-dimensional integrable 2-component viscous cosmology'',
    \GC {7} 301 (2001).

\bibitem{GavMel1}
       V.R. Gavrilov and V.N. Melnikov,
        ``Modern trends in multidimensional gravity and cosmology'',
         {\it in:\/} ``Advances in the Interplay Between Quantum and Gravity
             Physics'', P.G. Bergmann and V. de Sabbata (eds.),
             Kluwer Academic Publishers, 2002, p. 285-315.

\bibitem{DGavMel2}
        H. Dehnen, V.R. Gavrilov and V.N. Melnikov,
        ``General solutions for a flat Friedmann universe filled with a
    perfect fluid and a scalar field with an exponential potential'',
    \GC {8} 4 (32), 189 (2003).

\bibitem{GavMelAbd}
         V.R. Gavrilov, V.N. Melnikov and S.T. Abdyrakhmanov,
         ``Flat Friedmann Universe filled with dust and scalar field
         with multiple exponential potential'',
        \GRG {\bf 36}, 1579 (2004).

\bibitem{AlGavMel}
        J.-M. Alimi, V.R. Gavrilov and V.N. Melnikov,
        ``Multicomponent perfect fluid with
        variable parameters in $n$ Ricci-flat spaces'',
        Proceedings of ICGA6, 6-9 October 2003, Seoul, Korea,
    {\it J. Korean  Phys. Soc.} {\bf 44}, S148 (2004).

\bibitem{AlIvMel}
         J.-M. Alimi, V.D. Ivashchuk and V.N. Melnikov,
     ``Non-singular solutions in multidimensional model with scalar
     fields and exponential potential'',
     \GC {11} 1-2, 111 (2005).

\bibitem{MelGav}
    V.N. Melnikov and V.R. Gavrilov,
    ``2-component cosmological models with perfect fluid: exact solutions'',
     {\it In:\/} The Gravitational Constant: Generalized Gravitational
     Theories and Experiments, eds. V. de Sabbata, G.Gillies and V.N.
     Melnikov, Kluwer Academic Publishers, 2004, p. 247-268.

\bibitem{Cosm}
    M.A. Grebeniuk, V.D. Ivashchuk and V.N. Melnikov,
    ``Multidimensional cosmology for intersecting $p$-branes with
    static internal spaces'', \GC {4} 2(14), 145 (1998).

\bibitem{IM6I}
    V.D. Ivashchuk and V.N. Melnikov,
    {\it  Int. J. Mod. Phys. } {\bf D 4}, 167 (1995).

\bibitem{1}
    V.N. Melnikov and K.P. Staniukovich. In: ``Problems of Gravitation
     and Elementary Particle Theory (PGEPT)'', Atomizdat, Moscow, 1978,
    v. 9, p. 3 (in Russian).

\bibitem{2}
   N.A. Zaitsev and V.N. Melnikov, {\it in:\/} PGEPT, 1979, v. 10, p. 131.

\bibitem{7}
   V. Canuto et al. {\it Phys. Rev. } {\bf D 16}, 1643 (1977).

\bibitem{8}
   M. Novello and P. Rotelli, {\it J. Phys. A} {\bf 5}, 1488 (1972).

\bibitem{9}
   K.A. Bronnikov, V.N. Melnikov and K.P. Staniukovich, {\it prepr.\/}
   ITP-68-69, Kiev, 1968.

\bibitem{16}
   V.N. Melnikov. {\it Dokl. Acad. Nauk} {\bf 246}, N 6, 1351 (1979);\\
   V.N. Melnikov and S.V. Orlov, {\it Phys. Lett. } {\bf 70 A}, 4, 263
   (1979).

\bibitem{10}
   V.N. Melnikov and A.G. Radynov, {\it in:\/} PGEPT, 1984, v. 14, p. 73.

\bibitem{14}
  V.N. Melnikov and A.G. Radynov. In: ``On Relativity Theory'', Singapore,
  WS, 1985, v. 2, p. 196.

\bibitem{18}
  R. Hellings, {\it Phys. Rev. Lett. } {\bf 51}, 1609 (1983).

\bibitem{VP}
   D.A. Varshalovich and A.Y. Potekhin. {\it Space Sci. Rev. } {\bf 74}, 259
   (1995); {\it Pis'ma Atron. Zh.} {\bf 22}, 3 (1996);
   {\it Astron. Lett. } {\bf 22}, 6 (1996).

\bibitem{D}
   M. Drinkwater et al., astro-ph/9711290.

\bibitem{JP}
   J .D. Prestage et al., {\it Phys. Rev. Lett. } {\bf 74}, 3511 (1995).

\bibitem{STAB}
   K.A. Bronnikov and V.N. Melnikov, \GC {5} 306 (1999).

\bibitem{20}
   V.N. Melnikov et al., \GRG {17} 63 (1985).

\bibitem{19}
   V.N. Melnikov. In: ``Results of Science and Technology. Gravitation and
   Cosmology''. Ed.: V.N. Melnikov, 1991, v. 1, p. 49 (in Russian).

\bibitem{MT99}
  See all papers in the issue
  {\it Measurement Science and Technology}, {\bf 10}, No. 6 (1999).

\bibitem{MIWaseda}
   V.D. Ivashchuk and V.N. Melnikov, ``Problems of $G$ and
    multidimensional models'', {\it in:\/} Proc. JGRG 11, eds. J.
    Koga et al., Waseda Univ., Tokyo, 2002, pp. 405-409.

\bibitem{BMN-G}
   K.A. Bronnikov, V.N. Melnikov and M. Novello,`` Possible time
    variations of $G$ in scalar-tensor theories of gravity'',
    \GC {8} Suppl. II, 18--21 (2002).

\bibitem{M-G}
   V.N. Melnikov, ``Time variations of $G$ in different models'',
   Proc. 5th Int. Friedmann Seminar, Joao Pessoa, Brazil,
   {\it Int. J. Mod. Phys. } {\bf A 17}, 4325 (2002).

\bibitem{DIKM-G}
   H. Dehnen, V.D. Ivashchuk, S.A. Kononogov and V.N. Melnikov,
   ``On time variation of $G$ in multidimensional models with two
    curvatures'', \GC {11} 340 (2005).

\bibitem{P}
   E.V. Pitjeva, {\it in:\/} ``Dynamics and Astrometry of Natural and
   Artificial Celestial Bodies'', Kluwer Acad. Publ., Netherlands, 1997, p.
    251.

\bibitem{G}
   J.H. Gundlach and S.M. Merkowitz, {\it Phys. Rev. Lett.\/} {\bf 85}
   2869 (2000); gr-qc/0006043.

\bibitem{KolMel}
   N.I. Kolosnitsyn and V.N. Melnikov, ``Test of inverse square law
    through precession of orbits'', \GRG {36} 7, 1619--1624 (2004).

\bibitem{BKM-BW}
   K.A. Bronnikov, S.A. Kononogov and V.N. Melnikov,
    ``Brane world corrections to Newton's law'',
    \GRG {38} 1215--1232 (2006); gr-qc/0601114.

\bibitem{BKalpha}
   K.A. Bronnikov and S.A. Kononogov, ``Possible variations of the fine
   structure constant $\alpha$ and their metrological significance'',
    {\it Metrologia\/} {\bf 43}, 5, R1--R9 (2006).

\bibitem{Webb}
   J. Webb  et al., {\it Phys. Rev. Lett.\/} 87, 091301 (2001).

\bibitem{Mills}
    Ian M. Mills et al., {\it Metrologia} {\bf 43}, 227--246 (2006).

 \bibitem{m6}
    A. Ranada, {\it Europhys. Lett. } {\bf 63}, 653 (2002).

 \bibitem{m7}
    J.D. Anderson et al., \PRL {75} 3602 (1995); {\bf 81}, 2858 (1998).

\bibitem{m8}
    A. Das et al., \JMP {44} 5536 (2003).

\bibitem{m9}
    G. Modanese, \NPB {556} 397 (1999); gr-qc/9903085.

\bibitem{m10}
    D. \O stvang, \CQG {19} 4131 (2002); gr-qc/9910054.

\bibitem{m11}
    W.B. Belayev, gr-qc/9903016.

\bibitem{m12}
    R. Mansouri, F. Nasseri and M. Khorrami, \PLA {259} 194 (1999);
    gr-qc/9905052.

\bibitem{m13}
    J.W. Moffat, gr-qc/0405076.

\bibitem{m14}
    M. Milgrom, {\it Acta Phys. Pol. } {\bf B 32}, 3613 (2001).

\bibitem{m15}
    J. Bekenstein, \PRD {70} 083509 (2004); astro-ph/0403694.

\bibitem{m16}
    S. Calchi Novati et al., \GC {6} 173 (2000); astro-ph/0005104.

\bibitem{m17}
    J.-P. Mbelek and M. Lachieze-Rey, gr-qc/9910105.

\bibitem{m18}
       M.-T. Jaekel and S. Reynaud, {\it CQG} {22} 2135 (2005);
       {\it Mod. Phys. Lett. } {\bf A 20}, 1047 (2005).

\bibitem{m19}
    S. Turyshev, M. Nieto and J. Anderson, invited talk at the XXIIth
    Texas Symposium on Relativistic Astrophysics, Stanford University,
    December 13-17, 2004; gr-qc/0503021.

\bibitem{KKM}
    N.I. Kolosnitsyn, S.A. Kononogov and V.N. Melnikov.
    {\it Izmeritelnaya Technika}, 2007, No. 6,  p. 3.

\bibitem{KrKM}
    S.A. Kononogov, V.N. Melnikov and V.V. Khruschev,
    {\it Izmeritelnaya Technika}, 2007, No. 3, p. 3.

\bibitem{AIKM}
    J.-M. Alimi, V.D. Ivashchuk, S.A. Kononogov, V.N. Melnikov,
    ``Multidimensional cosmology with anisotropic fluid: acceleration
    and variation of $G$''. Proc. Int. Conf. on Gravitation, Cosmology,
    Astrophysics and Nonstationary Gas Dynamics, dedicated to Prof. K.P.
        Staniukovich's 90th birthday, Moscow, 2-6 March 2006,
    \GC {12} 173 (2006).

\bibitem{IKMN}
     V.D. Ivashchuk, S.A. Kononogov, V.N. Melnikov and  M. Novello,
    ``Non-singular solutions in multidimensional cosmology with a
    perfect fluid: acceleration and variation of $G$'',
    \GC {12} 273 (2006).

\bibitem{AIMAG}
     J.-M. Alimi, V.D. Ivashchuk and V.N. Melnikov, ``S-brane solution
    with acceleration and small enough variation of $G$'',
    \GC {13} 137 (2007) (this issue).

\bibitem{DIMParal}
     H. Dehnen, V.D. Ivashchuk and V.N. Melnikov, ``S-brane solutions
     with (anti-)self-dual parallel charge density form on a
     Ricci-flat submanifold'', \GC {13} 1 (49), 23 (2007).

\end{thebibliography}
\end{document}